\documentclass[12pt,preprint,prb]{revtex4}
\usepackage{graphicx}
\usepackage{psfrag}
\usepackage{amssymb,amsfonts,amsmath}

\newcommand{\lbe}[1]{\begin{equation} \label{#1}}
\newcommand{\ee}{\end{equation}}
\newcommand{\dd}{\mathrm{d}}
\newcommand{\e}{\mathrm{e}}
\newcommand{\ii}{\mathrm{i}}
\newcommand{\tr}{\mathrm{Tr}\,}

\begin{document}

\title{Progress in relativistic gravitational theory using the inverse scattering method}
\author{G. Neugebauer}
\author{R. Meinel} 
\affiliation{Theoretisch-Physikalisches Institut, Friedrich-Schiller-Universit\"at Jena, 
Max-Wien-Platz 1, 07743 Jena, 
Germany}
\email[e-mail: ]{neugebauer@tpi.uni-jena.de}

\begin{abstract}
The increasing interest in compact astrophysical objects (neutron stars,
binaries, galactic black holes) has stimulated the search for rigorous
methods, which allow a systematic general relativistic description of
such objects. This paper is meant to demonstrate the use of the inverse
scattering method, which allows, in particular cases, the treatment of
rotating body problems. The idea is to replace the investigation of the
matter region of a rotating body by the formulation of boundary values
along the surface of the body. In this way we construct solutions
describing rotating black holes and disks of dust (\lq\lq galaxies\rq\rq).
Physical properties of the solutions and consequences of the approach
are discussed. Among other things, the balance problem for two 
black holes can be tackled.
\end{abstract}
\maketitle

\section{Introduction}

The systematic investigation of neutron stars and binaries consisting of
pulsars and other compact objects  and the increasing evidence for the 
existence of (rotating) black holes have stimulated theoretical and numerical studies on rapidly 
rotating bodies in General Relativity. No doubt, realistic stellar models (e.g. neutron star models) 
require a careful physical analysis of their interior states and processes and, as a consequence, 
extensive {\em numerical} calculations. On the other hand, there is widespread interest for 
explicit solutions of the rotating body problem under simplifying assumptions. Such solutions could
provide a deeper insight into physical phenomena connected with spinning matter configurations and, 
moreover, serve as test beds for the numerical investigations mentioned before. A good example is the 
Kerr solution, which has enriched our knowledge of rotating black holes in an inestimable way.  
However, rigorous results for rotating bodies are relatively rare in General Relativity. Among 
other things, this is due to the mathematical difficulties with \lq free boundary value problems\rq, 
already known from Newton's gravitational theory, and to the specific complexity of the differential 
equations of Einstein's theory inside the body. Namely, the shape of the surface of a rotating 
self-gravitating fluid ball --- the best model for astrophysical
applications --- is a `compromise' 
between gravitational and centrifugal forces and not known {\em a priori}. (The surface is `free', 
i.e. not fixed from the very beginning.) Though there are powerful (soliton-) techniques to 
generate (formal) stationary axisymmetric solutions of Einstein's vacuum equations, no 
algorithm to integrate the interior field equations is available. Hence, at first glance, a 
boundary value description of rotating bodies seems to be questionable and inadequate. However, 
there are exceptional cases, in which the surface of the body has a known shape and the surface 
values provide enough information to construct the complete solution of the vacuum field equations. It is 
the intention of this paper to show that this is true for stationary black holes and disks of dust, 
which may be considered to be extremely flattened perfect fluid bodies. Moreover, it should become 
clear that our procedure, which is based on the inverse scattering method, opens an access to the 
not yet solved problem of the balance of two black holes and enables, in principle, the 
construction of black holes surrounded by dust rings (`AGN models').

Another interesting domain of application for the inverse scattering
method is colliding gravitational waves. This theory is out of the
scope of our paper. We refer to the article \cite{than} and references therein.

The present work is mainly based on the papers \cite{0} and \cite{3} but
it also contains substantial material not published before.

\section{The boundary value problem}

\begin{figure}
\psfrag{z}{\large{$\zeta$}}
\psfrag{r}{\large{$\rho$}}
\psfrag{iz}{\large{$\ii\bar{z}$}}
\psfrag{-iz}{\large{$-\ii z$}}
\psfrag{B}{\large{$\mathcal{B}$}}
\psfrag{C}{\large{$\mathcal{C}$}}
\psfrag{A+}{\large{$\mathcal{A}^+$}}
\psfrag{A-}{\large{$\mathcal{A}^-$}}
\psfrag{RK}{\large{$\Re K$}}
\psfrag{IK}{\large{$\Im K$}}
\psfrag{a}{\large{(a)}}
\psfrag{b}{\large{(b)}}
\centerline{\resizebox{12cm}{6.5cm}{\includegraphics{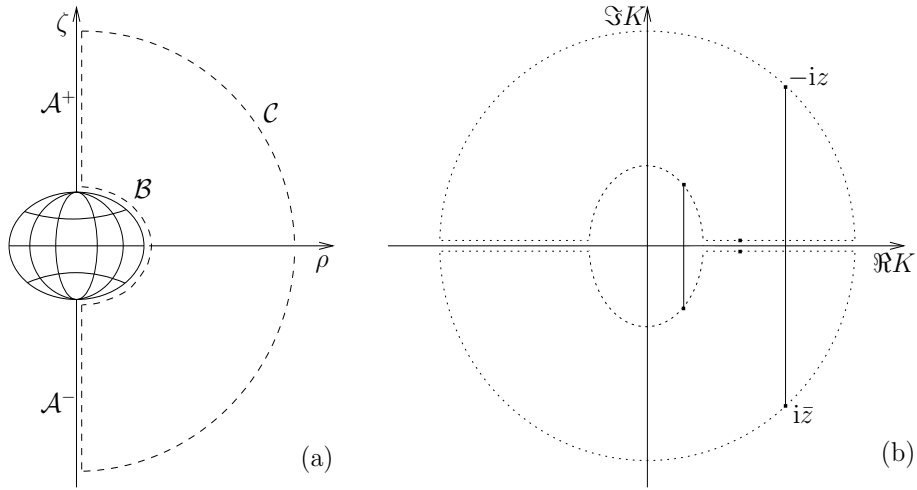}}}
 \caption{(a) A slice $\phi=\text{constant}, t=\text{constant}$, (b) 2-sheeted Riemann $K$-surface with branch points (dots) and two 
cuts (solid lines) \cite{0}}
 \label{fig1}
\end{figure}
We consider a simply connected axisymmetric and stationary body and describe its exterior vacuum 
gravitational field in Weyl-Lewis-Papapetrou coordinates
\lbe{2.1}
\dd
s^2=\e^{-2U}\left[\e^{2k}\left(\dd\rho^2+\dd\zeta^2\right)+\rho^2\dd\phi^2\right]-\e^{2U}\left(\dd
t+a\dd\phi\right)^2
\ee
where the `Newtonian' gravitational potential $U$ and the \lq
gravitomagnetic\rq\  potential $a$ are 
functions of  $\rho$ and $\zeta$ alone. Fig. 1(a) shows the boundaries of the vacuum region: 
$\cal{A^\pm}$ are the regular parts of the axis of symmetry $(\rho = 0)$, $\cal{B}$ is the 
surface of the body and $\cal{C}$ stands for spatial infinity. Later on, we will integrate along the 
dashed line and pick up information from the boundary values of the gravitational fields at 
$\cal{A^\pm}$, $\cal{B}$ and $\cal{C}$. The metric (\ref{2.1}) allows an Abelian group of 
motions $G_2$ with the generators (Killing vectors)
\lbe{2.2}
\begin{array}{rclrcll}
\xi^i&=&\delta^i_t,\qquad&\xi^i\xi_i&<&0\qquad&\mbox{(stationarity)}\\[1ex]
\eta^i&=&\delta^i_{\phi},\qquad&\eta^i\eta_i&>&0\qquad&\mbox{(axisymmetry)}
\end{array}
\ee
where the Kronecker symbols $\delta^i_t$ and $\delta^i_{\phi}$ indicate that $\xi^i$ has only a $t$-component 
whereas 
$\eta^i$ points in the azimuthal $\phi$-direction (its trajectories are closed circles!). Obviously, 
\lbe{2.3}
\e^{2U}=-\xi^i\xi_i,\qquad a=-\e^{-2U}\eta_i\xi^i
\ee
is a coordinate-free representation of the two relativistic
gravitational fields $U$ (generalization of the Newtonian gravitational
potential) and $a$ (gravitomagnetic potential). To 
get a unique definition of $U$ and $a$, we prescribe their behaviour at infinity. Assuming that the 
space-time has to be flat at large distances from the body and can be described by a Minkowskian 
line element (\ref{2.1}) in cylindrical coordinates, we are led to the boundary conditions 
\lbe{2.4}
\mathcal{C}:\quad U\to 0,\quad a\to 0,\quad k \to 0
\ee
Any linear transformation 
\begin{equation}\label{2.5}
t'=t,\quad\phi'=\phi-\omega t
\end{equation}
introduces a frame of reference which rotates with a 
constant angular velocity $\omega$ with respect to that asymptotic
Minkowski space.

To describe stationarity and axixymmetry in that rotating system one would use the Killing
vectors
\begin{equation}\label{2.5a}
\tilde\xi^i=\xi^{i'}+\omega\eta^{i'},\quad \tilde\eta^i=\eta^{i'},
\end{equation} 
instead of \eqref{2.2}.

Regularity of the metric along ${\mathcal{A}^{\pm}}$ means
\lbe{2.6}
\mathcal{A}^{\pm}:\quad a=0,\quad k=0.
\ee
These conditions express the fact that ${\mathcal{A}^{\pm}}$ is an axis of symmetry ($a=0$) and 
ensure elementary flatness along the axis ($k=0$). The behaviour of $U$ and $a$ at the surface 
$\mathcal{B}$ of the body depends on the physical nature of it. Rotating
{\em perfect fluids} are 
characterized by a {\em four-velocity field} $u^i$ consisting of a linear combination of the two Killing 
vectors,
\lbe{2.7}
u^i=\e^{-V}\left(\xi^i+\Omega\eta^i\right) \quad , u^iu_i=-1
\ee
where $\Omega$ is the angular velocity of the body, and an {\em invariant
scalar pressure} $p$,
which is, for rigid rotation,
\begin{equation}\label{2.7a}
\Omega=\Omega_0,\quad(\Omega_0\ \textrm{a constant})
\end{equation}
a function of $V$ alone,
\lbe{2.8}
p=p(V),
\ee
as a consequence of the Euler equations.
Along the surface of the body (if it exists) the pressure has to vanish, \newline
$p(V_0)=0$, i. e. $V$ must be a constant along $\mathcal{B}$,
\lbe{2.9}
\mathcal{B}:\quad \e^{2V}\equiv -(\xi^i+\Omega_0\eta^i)(\xi_i+\Omega_0\eta_i)=\e^{2V_0}
\ee
That is a further boundary condition. When identifying $\omega$ in (\ref{2.5}) and $\Omega_0$ we 
introduce a frame of reference co-rotating with the body and may interprete $V$ as the co-rotating 
`Newtonian' potential, cf. (\ref{2.3}) and (\ref{2.9}). Interestingly,
the event horizon $\mathcal{H}$
of a stationary (axisymmetric) black hole 
behaves like an `ordinary' perfect fluid surface (\ref{2.9}). Namely, one can show that a 
linear combination of the two Killing vectors, $\xi^i+\Omega_\text{H} \eta^i$ has a vanishing norm along 
$\mathcal{H}$,
\lbe{2.10}
\mathcal{H}:\quad \e^{2V}\equiv(\xi^i+\Omega_\text{H}\eta^i)(\xi_i+\Omega_\text{H}\eta_i)=0,
\ee
where $\Omega_\text{H}$ is the angular velocity of the horizon. Hence we may include black holes in our 
scheme, see Fig. 1(a), for $V_0\to -\infty$ and $\mathcal{H}\equiv \mathcal{B}$. It will turn out that 
(\ref{2.10}) together with the 
correct positioning of the horizon $\mathcal{H}$ in Weyl-Lewis-Papapetrou coordinates together with the 
asymptotic behaviour (\ref{2.4}) of the (invariant) potentials (\ref{2.3}) suffices for an explicit 
construction of the Kerr solution - thus providing a simple constructive uniqueness proof for stationary 
axisymmetric black holes. 
On the other hand, the condition \eqref{2.9} is not sufficient to calculate the gravitational 
vacuum field of rotating {\em perfect fluid balls}. However, in the disk of dust limit of such fluid 
configurations the field equations themselves will provide the missing boundary condition along the 
surface $\mathcal{B}$ of the disk, see \cite{3}. Starting with that completed set of boundary 
conditions we will be able to construct the global solution for the rigidly rotating disk of dust.

\section{The field equations}

The vacuum Einstein equations for the metric coefficients $k$, $U$, $a$ are equivalent to the Ernst 
equation
\lbe{2.11}
(\Re f)\left(f_{,\rho\rho}+f_{,\zeta\zeta}+\frac{1}{\rho}f_{,\rho}\right)=f_{,\rho}^2 + f_{,\zeta}^2
\ee
for the complex function 
\begin{equation}\label{2.11a}
f(\rho,\zeta)=\e^{2U}+\ii b,
\ee
where $b$ replaces $a$ via 
\lbe{2.12}
a_{,\rho}=\rho \e^{-4U}b_{,\zeta},\quad a_{,\zeta}=-\rho \e^{-4U}b_{,\rho}
\ee
and $k$ can be calculated from
\lbe{2.13}
k_{,\rho}=\rho\left[U_{,\rho}^2-U_{,\zeta}^2+\frac{1}{4}\e^{-4U}\left(b_{,\rho}^2-b_{,\zeta}^2\right)\right],\quad
k_{,\zeta}=2\rho\left[U_{,\rho}U_{,\zeta}+\frac{1}{4} \e^{-4U}b_{,\rho}b_{,\zeta}\right].
\ee
As a consequence of the Ernst equation (\ref{2.11}), the integrability conditions $a_{,\rho \zeta}=a_{,\zeta 
\rho}$ and $k_{,\rho \zeta}=k_{,\zeta \rho}$ 
are automatically satisfied such that the metric functions $a$ and $k$ may be calculated via line 
integration from the Ernst potential $f$. Thus, it is sufficient to discuss the Ernst equation 
alone.

\section{The Linear Problem}

The existence of a Linear Problem (LP) for the Ernst equation
\cite{4,5,6,7,8,9} is the corner stone of our analysis 
since 
it provides a suitable instrument for tackling boundary value problems: the inverse scattering method (ISM). 
Here we 
will use a 
`local' version \cite{10} of the Linear Problem, 
\begin{equation}\label{2.14}
 \begin{aligned}
\mathbf{\Phi}_{,z}=&\, \left\{\left(\begin{array}{rr} B&0\\0&A\end{array}\right)+\lambda\left(\begin{array}{rr} 
0&B\\A&0\end{array}\right) \right\} \mathbf{\Phi}, \\[2.7ex]
\mathbf{\Phi}_{,\bar{z}}=&\, \left\{\left(\begin{array}{rr} \bar{A}&0\\0&\bar{B}\end{array}\right)+ 
\displaystyle\frac{1}{\lambda}\left(\begin{array}{rr}
0&\bar{A}\\\bar{B}&0\end{array}\right) \right\} \mathbf{\Phi},
 \end{aligned}
\end{equation}
where $\mathbf{\Phi}(z,\bar{z},\lambda)$ is a $2\times 2$ matrix depending on the spectral parameter
\lbe{2.15}
\lambda=\sqrt{\frac{K-\ii\bar{z}}{K+\ii z}}
\ee
as well as on the complex coordinates $z=\rho+\ii\zeta$, $\bar{z}=\rho-\ii\zeta$, whereas $A, B$ and the complex 
conjugate quantities 
$\bar{A}$, $\bar{B}$ are 
functions of $z,\bar{z}$ (or $\rho,\zeta$) and do not depend on $K$. From the integrability condition and the formulae
\lbe{2.16}
\lambda_{,z}=\frac{\lambda}{4\rho}\left(\lambda^2-1\right),\quad \lambda_{,\bar{z}}=\frac{1}{4 \rho 
\lambda}\left(\lambda^2-1\right)
\ee
it follows that a matrix polynomial in $\lambda$ has to vanish. This yields the set of first order 
differential 
equations
\lbe{2.17}
A_{,\bar{z}}=A(\bar{B}-\bar{A})-\frac{1}{4\rho}(A+\bar{B}),\quad 
B_{,\bar{z}}=B(\bar{A}-\bar{B})-\frac{1}{4\rho}(B+\bar{A}).
\ee
The system has the `first integrals'  
\lbe{2.18}
A=\frac{f_{,z}}{f+\bar{f}},\quad B=\frac{\bar{f}_{,z}}{f+\bar{f}}.
\ee
Resubstituting $A$ and $B$ in the equations \eqref{2.17} one obtains the Ernst equation (\ref{2.11}). Thus, 
the 
Ernst 
equation is the integrability condition of the LP (\ref{2.14}). Vice versa, if $f$ is a solution to the Ernst 
equation, the 
matrix $\mathbf{\Phi}$ calculated from (\ref{2.14}) does not depend on the path of integration. The idea of 
the inverse scattering method (ISM) 
is to discuss $\mathbf{\Phi}$, for fixed but arbitrary values of $z,\bar{z}$, as a holomorphic function of 
$\lambda$ (or $K)$ 
and to 
calculate $A,B$ and finally $f$ afterwards. To obtain the desired information about the holomorphic structure in $\lambda$, we 
will integrate the Linear System along the dashed line in Fig. 1(b) making use of the conditions (\ref{2.4}), 
(\ref{2.6}), 
(\ref{2.9}) or (\ref{2.10}). In this way, we will solve the {\em direct problem} of the ISM and obtain 
$\mathbf{\Phi}(z,\bar{z},\lambda)$ 
for $z,\bar{z} \in \mathcal{A}^{\pm}, \mathcal{B}, \mathcal{C}$. 
It turns out that the holomorphic structure remains unchanged by an
extension of $z,\bar{z}$ off the axis of symmetry into 
the entire
vacuum region 
such that one can construct functions $\mathbf{\Phi}$ with prescribed properties in $\lambda$ from which one 
obtains the 
desired 
solution $f(z,\bar{z})$ everywhere in the vacuum region. This second step can be very technical and will, in 
general, 
lead to linear 
integral equations for $\mathbf{\Phi}$. In some circumstances, $\lambda$ may be replaced by $K$. For this purpose,
it may 
be helpful to 
discuss the mapping (\ref{2.15}) of the two-sheeted Riemann surface of $K$ onto the $\lambda$-plane for different 
values of $\rho, \zeta$ (or equivalently $z,\bar{z}$). 
Fig. 1(b) shows the position of the branch points
$K_\text{B}=\ii\bar{z},\bar{K}_\text{B}=-\ii z$ for the marked path 
$\mathcal{A}^+\mathcal{C}\mathcal{A}^-\mathcal{B}$ of Fig. 1(a). It reflects the 
 slice $\phi=\text{constant},t=\text{constant}$ (Fig. \ref{fig1}(a)) and
 indicates, in particular, the position and shape of the body. Note that
 $\mathbf\Phi$ is not defined in the non-vacuum domain inside the
 circular contour around the origin.

Consider now a Riemann surface with confluent branch points $K_\text{B}=\bar{K}_\text{B}=\zeta \in 
\mathcal{A}^+$. Here $\lambda$ degenerates and takes 
the values $\lambda=-1$ for $K$'s in the lower sheet, say, and $\lambda=+1$ for $K$'s in the upper sheet 
$(K\ne K_\text{B})$.

We will now travel along the dashed line of Fig. 1(a) starting from and returning to any point $\rho=0$, $\zeta \in 
\mathcal{A}^+$. (In Fig. 1(b) this corresponds to the bold faced points on the
real axis.) Note that 
$\lambda=-1$ for all 
$K$'s $(K\ne \zeta)$ in the lower and $\lambda=+1$ for all $K$'s  $(K\ne \zeta)$ in the upper sheet of the Riemann 
$K$-surface belonging to 
axis values $\rho=0$, $\zeta \in \mathcal{A}^{\pm}$ (the corresponding branch points cling to
either side of the real axis in Fig. 1(b)). For $\rho, \zeta \in 
\mathcal{C}$, the cut between the branch points (e.g., right solid 
line in Fig. 
1(b)) points over the entire $K$-surface and puts `upper' $K$ values into the lower sheet and `lower' $K$ values 
into the `upper' 
sheet. As a consequence, $\lambda$ will change from $\pm 1$ to $\mp 1$ between $\rho = 0$, $\zeta=+\infty$ and 
$\rho=0$, $\zeta=-\infty$ \cite{11}.  
This `exchange of 
sheets' is important for the solution of the linear problem: The initial value 
$\mathbf{\Phi}(\rho_0,\zeta_0,\lambda)$ can (and must) be 
fixed only in 
{\em one} sheet of the $K$-surface. The dependence on $K$ in the other
sheet follows by integration of the LP (\ref{2.14}) along a 
suitable path \cite{11}.

We will divide the integration of the LP \eqref{2.14} along the closed
dashed line of Fig. \ref{fig1}(a) into two steps:
\begin{enumerate}
\item[(i)] Integrating along
$\mathcal{A}^+\mathcal{C}\mathcal{A}^-$ \mbox{}\\
This step can be performed without particular knowledge about the body
and leads to a \lq\lq\emph{general} solution\rq\rq\ for $\mathbf{\Phi}$
on the regular parts $\mathcal{A}^\pm$ of the symmetry axis.
\item[(ii)] Integrating along $\mathcal{B}$\mbox{}\\
Here we confine ourselves to black holes and disks of dust.
\end{enumerate}

\section{Solution of the direct problem}

\subsection{Axis and Infinity}

Without loss of generality the matrix $\mathbf{\Phi}$ may be assumed to have the structure
\lbe{5.1}
\mathbf{\Phi}=\left(\begin{array}{rr}\psi(\rho,\zeta,\lambda)&\psi(\rho,\zeta,-\lambda)\\[1ex] 
\chi(\rho,\zeta,\lambda)&-\chi(\rho,\zeta,-\lambda)\end{array}\right)
\ee
together with 
\lbe{5.2}
\overline{\psi\left(\rho,\zeta,\frac{1}{\bar{\lambda}}\right)}=\chi(\rho,\zeta,\lambda).
\ee
Note that both columns of $\mathbf\Phi$ are independent solutions of
\eqref{2.14}. The particular form of \eqref{5.1} is equivalent to
\begin{equation}\label{5.2a}
\mathbf\Phi(-\lambda)=\left(\begin{matrix}1&0\\0&-1\end{matrix}\right)\mathbf\Phi(\lambda)\left(\begin{matrix}0&1\\1&0\end{matrix}\right).
\end{equation}
For $K\to \infty$ and $\lambda=-1$ the functions $\psi,\chi$ may be normalized by
\lbe{5.3}
\psi(\rho,\zeta,-1)=\chi(\rho,\zeta,-1)=1
\ee
Finally, the solution to the Ernst equation can be read off at $\lambda=1$ $(K\to \infty)$,
\lbe{5.4}
f(\rho,\zeta)=\chi(\rho,\zeta,1),\quad \left(\overline{f(\rho,\zeta)}=\psi(\rho,\zeta,1)\right).
\ee
Remarkably enough, the Ernst equation retains its form in the frame of reference co-rotating with the body 
($\omega=\Omega_0$). 
This is a consequence of (\ref{2.3}) and \eqref{2.5a} and implies the existence of a Linear Problem 
(\ref{2.14}) in 
the co-rotating system. In particular, the $\mathbf{\Phi}$-matrices of both systems of reference are connected 
by the relation 
\lbe{5.5}
\begin{array}{rcl}
\mathbf{\Phi}'&=&\Bigg[\left(\begin{array}{cc}1+\Omega_0 a-\Omega_0\rho \e^{-2U}&0\\[3ex]0&1+\Omega_0 a+\Omega_0\rho \e^{-2U}\end{array}\right)\\[5ex]
&&\quad\qquad+\ii(K+\ii z)\Omega_0\e^{-2U}\left(\begin{array}{rr}-1&
-\lambda\\[1ex]\lambda&1\end{array}\right)\Bigg]
\mathbf{\Phi}.
\end{array}
\ee
Henceforth, a prime marks \lq co-rotating\rq\  quantities.
We can now realise our programme and integrate the Linear Problem
(\ref{2.14}) along the part $\mathcal{A}^+\mathcal{C}\mathcal{A}^-$ of
the dashed line in Fig. 1(a). 
Using \eqref{2.14} along $\mathcal{A}^\pm$ and \eqref{2.18} one finds
for the axis values of $\mathbf{\Phi}$
\lbe{5.6}
\mathcal{A}^+:\quad \mathbf{\Phi} 
=\left(\begin{array}{rr}\overline{f(\zeta)}&1\\[1ex]f(\zeta)&-1\end{array}\right)\left(\begin{array}{rr}F(K)&0
\\[1ex]G(K)&1\end{array}\right)
\ee
\lbe{5.7}
\mathcal{A}^-:\quad \mathbf{\Phi} 
=\left(\begin{array}{rr}\overline{f(\zeta)}&1\\[1ex]f(\zeta)&-1\end{array}\right)\left(\begin{array}{rr}1&G(K)
\\[1ex] 0&F(K)\end{array}\right),
\ee
where $f(\zeta) = f (\rho=0, \zeta)$ is the axis value of the Ernst potential and $F(K), G(K)$ are integration 
`constants' 
depending on $K$ alone. The particular form of (\ref{5.6}) is due to the initial condition $\psi=\chi=1$ for 
some 
$\rho_0=0,\zeta=\zeta_0\in \mathcal{A}^+$, $\lambda=-1$ ($K$ in the 
lower sheet), which  fixes the second column of $\mathbf{\Phi}$ in (\ref{5.6}), cf. (\ref{5.1}). 
The first column corresponds to the upper ($\lambda=1$) sheet and
represents a general integral with the two integration \lq constants\rq\
$F(K), G(K)$ which cannot be specified here. Along $\mathcal{C}$,
$\mathbf{\Phi}=\mathbf{\Phi}(K)$ does not depend on $\rho$ and $\zeta$,
since $A$ and $B$ vanish, cf. \eqref{2.18}. The \lq exchange of
sheets\rq\ along $\mathcal{C}$, see Fig. \ref{fig1}(b),
together with \eqref{5.2a} leads to the particular form of $\mathbf\Phi$
on $\mathcal A^-$. The representations \eqref{5.6}, \eqref{5.7} describe
the behaviour of $\psi$ and $\chi$ in both sheets. Nevertheless, one may
wish to consider the matrix $\mathbf\Phi$ as a  whole as a unique function
of $\lambda$, which is therefore defined on both sheets of the
$K$-surface. From this point of view, the equations \eqref{5.6},
\eqref{5.7} describe $\mathbf\Phi$ on one sheet only (say, on the upper
sheet). Its values on the other (lower) sheet follow from \eqref{5.2a}.

Combining (\ref{5.6}),(\ref{5.7}) with (\ref{5.5}) we obtain the axis 
values 
in the 
co-rotating system 
\lbe{5.8}
\begin{array}{rcl}
\mathcal{A}^+:\quad 
\mathbf{\Phi}'&=&\left[\mathbf{1}+\ii (K-\zeta)\Omega_0\e^{-2U}\left(\begin{array}{rr}-1&-1\\1&1\end{array}\right)
\right]\times\\[3ex]
&&\qquad\qquad\qquad\qquad\times\left[\left(\begin{array}{rr}\overline{f(\zeta)}&1\\f(\zeta)&-1\end{array}\right)\left(\begin{array}{rr}F(K)&0\\G(K)&1\end{array}\right)\right]
\end{array}
\ee
\lbe{5.9}
\begin{array}{rcl}
\mathcal{A}^-:\quad 
\mathbf{\Phi}'&=&\left[\mathbf{1}+\ii (K-\zeta)\Omega_0\e^{-2U}\left(\begin{array}{rr}-1&-1\\1&1\end{array}\right)
\right]\times\\[3ex]
&&\qquad\qquad\qquad\qquad\times\left[\left(\begin{array}{rr}\overline{f(\zeta)}&1\\f(\zeta)&-1\end{array}\right)\left(\begin{array}{rr}1&G(K)\\0&F(K)\end{array}\right)\right],
\end{array}
\ee
where $\mathbf{1}$ ist the $2\times 2$ unit matrix.
At the branch points $K_\text{B}=\zeta$ of $K$-surfaces belonging to axis values $\rho=0, \zeta \in 
\mathcal{A}^{\pm}$, $\psi$ and $\chi$ must be unique, i.e.
\lbe{5.10}
\mathcal{A}^+(K_\text{B}=\zeta):\quad \mathbf{\Phi}=\left(\begin{array}{rr}\psi&\psi\\ 
\chi&-\chi\end{array}\right)=\left(\begin{array}{rr}\overline{f(\zeta)}&1\\f(\zeta)&-1\end{array}\right)\left(
\begin{array}{rr}F(\zeta)&0\\G(\zeta)&1\end{array}\right)
\ee
\lbe{5.11}
\mathcal{A}^-(K_\text{B}=\zeta):\quad \mathbf{\Phi}=\left(\begin{array}{rr}\psi&\psi\\ 
\chi&-\chi\end{array}\right)=\left(\begin{array}{rr}\overline{f(\zeta)}&1\\f(\zeta)&-1\end{array}\right)\left
(\begin{array}{rr}1&G(\zeta)\\0&F(\zeta)\end{array}\right)
\ee
whence
\lbe{5.12}
\mathcal{A}^+:\quad F(\zeta)=\frac{2}{f(\zeta)+\overline{f(\zeta)}},\quad 
G(\zeta)=\frac{f(\zeta)-\overline{f(\zeta)}}{f(\zeta)+\overline{f(\zeta)}}
\ee

\lbe{5.13}
\mathcal{A}^-:\quad F(\zeta)=\frac{2f(\zeta)\overline{f(\zeta)}}{\overline{f(\zeta)}+f(\zeta)},\quad 
G(\zeta)=\frac{\overline{f(\zeta)}-f(\zeta)}{f(\zeta)+\overline{f(\zeta)}}
\ee
Thus, $F(K)$ and $G(K)$ consist in a unique way of analytic
continuations of the real and imaginary parts of the 
axis values of the Ernst potential $f(\zeta)$. Vice versa, 
$f(\zeta)$ follows from $F(K), G(K)$ for $K=\zeta$.
Interestingly, the determinants of $\mathbf{\Phi}$ and $\mathbf{\Phi}'$ can be expressed in terms of $\Re 
f$, $\Re f'$ and 
$F(K)$. 
From (\ref{2.14}) ($\mathrm{Tr\ } \mathbf{\Phi}_{,z}\mathbf{\Phi}^{-1}=(\ln \det \mathbf{\Phi})_{,z}$) , (\ref{2.18})
and (\ref{5.6})--(\ref{5.9}), we have
\lbe{5.14}
\det \mathbf{\Phi} = -2\e^{2U} F(K),\qquad \det \mathbf{\Phi}' = -2\e^{2V}F(K)
\ee
where $\e^{2U}=\Re f$ and $\e^{2V} = \Re f'$ $(U=U(\rho,\zeta),
V=V(\rho,\zeta))$.

We may now interpret the result of \eqref{5.8}--\eqref{5.13} of the
integration of the LP along $\mathcal{A}^+\mathcal{C}\mathcal{A}^-$: On the
regular parts $\mathcal{A}^\pm$ of the symmetry axis, $\mathbf{\Phi}$
and $\mathbf{\Phi}'$ can explicitly be expressed in terms of the axis
values $f(\zeta)$ of the Ernst potential and its analytic continuations
$F(K), G(K)$. To calculate $f(\zeta)$ one needs boundary values on
$\mathcal{B}$. Accordingly, the integration along $\mathcal{B}$ depends
on the physical nature of the rotating body and can be performed in
particular cases only. In the next section we will discuss black holes
and rigidly rotating disks of dust.

\subsection{Surface}

\subsubsection{One black hole}\label{5.B.1}
We identify the surface $\mathcal{B}$ with the horizon $\mathcal{H}$.
In Weyl coordinates, the event horizon $\mathcal{H}$ of a single black hole covers the domain \cite{12},  
\lbe{5.15}
\mathcal{H}:\quad \rho=0,\quad K_1\ge \zeta\ge K_2.
\ee
(In Fig. \ref{fig1}(a), the surface $\mathcal{B}$ degenerates to a `straight line' connecting the regular parts $\mathcal{A^-, A^+}$ of the 
axis of symmetry.) Along $\mathcal{H}$, $\e^{2V}$ has to vanish, see
(\ref{2.10}),
\begin{equation}\label{5.15a}
\mathcal{H}:\quad \e^{2V}\equiv(\xi^i+\Omega_0\eta^i)(\xi_i+\Omega_0\eta_i)=0\quad(\Omega_0=\Omega_\text{H}).
\end{equation}
 Because of 
\lbe{5.16}
\e^{2V}=\e^{2U} \left(\left[1+\Omega_0a\right]^2 - \Omega_0^2\rho^2
\e^{-4U}\right),
\ee
cf. (\ref{2.3}), this implies
\lbe{5.17}
\mathcal{H}:\quad 1+\Omega_0a=0.
\ee
$\mathbf{\Phi}$ and $\mathbf{\Phi}'$ can now be calculated along the horizon $\mathcal{H}$. From (\ref{2.14}), (\ref{5.15}),\eqref{5.15a}, (\ref{5.17}) and (\ref{5.5}) we obtain 
\lbe{5.18}
\mathcal{H}:\begin{array}{rcr}\quad \mathbf{\Phi}&=&\left(\begin{array}{rr}\overline{f(\zeta)}&1\\[1ex] f(\zeta)&-1\end{array}\right) \left(\begin{array}{rr}U(K)&V(K)\\[1ex]W(K)&X(K)\end{array}\right),\\[3ex] \mathbf{\Phi}'&=&2\ii\Omega_0(K-\zeta) \left(\begin{array}{rr}-1&0\\[1ex]1&0\end{array}\right)\left(\begin{array}{rr}U(K)&V(K)\\[1ex]W(K)&X(K)\end{array}\right).\end{array} 
\ee
The Ernst equations have to hold at $K_1$ and $K_2$ too. Hence, $\mathbf{\Phi}$ and $\mathbf{\Phi}'$ must be continuous in $K_1$ and $K_2$. 
Considering (\ref{5.6})--(\ref{5.9}) and \eqref{5.18}, we are led to the conditions
\lbe{5.19}
\begin{array}{rcl}
\left(\begin{array}{cr}f_1&-1\\[1ex]f_1+2\ii\Omega_0(K-K_1)&-1\end{array}\right)\left(\begin{array}{rr}F&0\\[1ex]G&1\end{array}\right)&=&\\[3ex]
&&\hspace{-3cm}\left(\begin{array}{cr}f_1&-1\\[1ex]2\ii\Omega_0(K-K_1)&0\end{array}\right)\left(\begin{array}{rr}U&V\\[1ex]W&X\end{array}\right),\\[7ex]
\left(\begin{array}{cr}f_2&-1\\[1ex]f_2+2\ii\Omega_0(K-K_2)&-1\end{array}\right)\left(\begin{array}{rr}1&G\\[1ex]0&F\end{array}\right)&=&\\[3ex]
&&\hspace{-3cm}\left(\begin{array}{cr}f_2&-1\\[1ex]2\ii\Omega_0(K-K_2)&0\end{array}\right)\left(\begin{array}{rr}U&V\\[1ex]W&X\end{array}\right),
\end{array}
\ee
where $f_1 = f(\zeta = K_1)$ and $f_2 = f(\zeta = K_2)$. Note that $f_1$
and $f_2$ are imaginary, see (\ref{5.14}).

Eliminating the $UVWX$ matrix, we obtain
\begin{equation}\label{5.19a}
\mathcal{N}=\left(\mathbf{1}+\frac{\mathbf{F}_1}{2\ii\Omega_0(K-K_1)}\right)\left(\mathbf{1}+\frac{\mathbf{F}_2}{2\ii\Omega_0(K-K_2)}\right),
\end{equation}
where
\begin{equation}\label{5.19b}
\mathbf{F}_1=\left(\begin{matrix} -f_1 & 1\\ -f_1^2 &
f_1\end{matrix}\right),
\quad\mathbf{F}_2=\left(\begin{matrix} f_2 & -1\\ f_2^2 & -f_2\end{matrix}\right),
\end{equation}
\begin{equation}\label{5.19c}
\mathcal{N}=\left(\begin{matrix}F & -G \\ G & (1-G^2)/F\end{matrix}\right).
\end{equation}
Obviously, the elements of $\mathcal{N}$ are regular everywhere in the
complex $K$-plane with the exception of the two simple poles at
$K_1$ and $K_2$ ($\Im K_1=0=\Im K_2$). The sum of the off-diagonal
elements in \eqref{5.19c} must be zero. This requirement leads to the
constraints
\lbe{5.20}
f_1=-f_2,\qquad \Omega_0=\frac{\ii f_1(1+f_1^2)}{(K_1-K_2)(1-f_1^2)}.
\ee
$F(K)$ and $G(K)$ take the form
\lbe{5.21}
F(K)=\frac{4\Omega_0^2(K^2-K_1^2)+4\ii\Omega_0 f_1
K-2f_1^2}{4\Omega_0^2(K^2-K_1^2)},\qquad G(K)=\frac{4\ii\Omega_0 K_1+2f_1}{4\Omega_0^2(K^2-K_1^2)}.
\ee
Here we have chosen $K_1 = -K_2$, i.e., we have set the horizon in a
symmetric position in the $\rho,\zeta$-plane. Making use of \eqref{5.12}
and \eqref{5.13} and eliminating $\Omega_0$ by the second constraint
equation we obtain the axis potential
\lbe{5.22}
\mathcal{A}^+: \quad f=\frac{\zeta(1+f_1^2)+(f_1^2-1+2f_1)K_1}{\zeta(1+f_1^2)+(1-f_1^2+2f_1)K_1}.
\ee

It can be useful to introduce  the multipole moments mass $M$ and
angular momentum~$J$ by an asymptotic expansion of $f$,
\begin{equation}\label{5.221}
M=\frac{1-f_1^2}{1+f_1^2}K_1,\qquad\frac{J}{M}=\alpha=\frac{2\ii f_1 K_1}{1+f_1^2}
\end{equation}
and to replace $f_1, K_1$ in \eqref{5.21},\eqref{5.22} and \eqref{5.20}:
\begin{equation}\label{5.222}
F(K)=\frac{(K+M)^2+\alpha^2}{K^2+\alpha^2-M^2},\qquad G(K)=\frac{2\ii
M\alpha}{K^2+\alpha^2-M^2}\ .
\end{equation}

To represent $f(\zeta)$, a simplifying parameterization is advisible,
\begin{equation}\label{5.223}
f_1=\ii\tan\varphi/2,\quad \alpha=-M\sin\varphi,\quad
K_1=-K_2=\sqrt{M^2-\alpha^2}=M\cos\varphi,\quad \varphi=\overline\varphi.
\end{equation}
This yields
\begin{equation}\label{5.224}
\mathcal{A}^+:\quad f=\frac{(\zeta-M)+\ii M\sin\varphi}{(\zeta+M)+\ii M\sin\varphi}.
\end{equation}

Finally, the second constraint equation \eqref{5.20} becomes the
well-known equation of state of black hole thermodynamics,
\begin{equation}\label{5.225}
2M\Omega_0=\frac{M}{\alpha}-\sqrt{\frac{M^2}{\alpha^2}-1},
\end{equation}
connecting the angular velocity of the horizon with mass and angular momentum.

\subsubsection{Two aligned black holes}

The same procedure can be used to tackle the balance problem for two
black holes. The question is whether the spin-spin repulsion of two
aligned stationary black holes can compensate their gravitational
attraction.

Here we have two horizons $\mathcal{H}_1$ and $\mathcal{H}_2$
\begin{equation}\label{5.22a}
\mathcal{H}_1:\quad\rho=0,\  K_1\ge \zeta \ge K_2,\quad\mathcal{H}_2:
\rho=0,\  K_3\ge \zeta\ge K_4
\end{equation}
separated by a piece of the regular symmetry axis $\mathcal{A}^0$
\begin{equation}\label{5.22b}
\mathcal{A}^0:\quad K_2\ge\zeta\ge K_3.
\end{equation}

As a characteristic black hole property, the norm of the Killing vectors
of the co-rotating frameworks has to vanish along the horizons,
\begin{equation}\label{5.22c}
\mathcal{H}_1:\quad
(\xi_i+\Omega_0^1\eta_i)(\xi^i+\Omega_0^1\eta^i)=0,\qquad
\mathcal{H}_2:\quad (\xi_i+\Omega_0^2\eta_i)(\xi^i+\Omega_0^2\eta^i)=0,
\end{equation}
where $\Omega_0^1,\Omega_0^2$ are the constant angular velocities of the
respective horizons.

Following the arguments for one black hole, we arrive at \cite{14}
\begin{equation}\label{5.22d}
\mathcal{N}=\prod_{i=1}^4\left(\mathbf{1}+\frac{\mathbf{F}_i}{2\ii\Omega_i(K-K_i)}\right)
\end{equation}
where $\Omega_1=\Omega_2=\Omega_0^2,\ \Omega_3=\Omega_4=\Omega_0^2$ and
\begin{equation}\label{5.22e}
\mathbf{F}_i=(-1)^i\left(\begin{matrix} f_i & -1\\f_i^2 & -f_i\end{matrix}\right),
\end{equation}
whence
\begin{equation}\label{5.22f}
\begin{aligned}
F(K) = &\  \frac{p_4(K)}{(K-K_1)(K-K_2)(K-K_3)(K-K_4)}\\ G(K) 
=  &\ \frac{p_2(K)}{(K-K_1)(K-K_2)(K-K_3)(K-K_4)},
\end{aligned}
\end{equation}
where $p_4(K)$ and $p_2(K)$ are polynomials in $K$ of the indicated
orders. From \eqref{5.22f} together with \eqref{5.10}, \eqref{5.11} we may read off
the axis values of the Ernst potential. For the upper axis we obtain the
structure
\begin{equation}\label{5.22g}
\mathcal{A}^+:\quad f(\zeta)=\frac{q_2(\zeta)}{Q_2(\zeta)}.
\end{equation}
We need not use the representation of the explicit form of the second
order polynomials $q_2(\zeta),Q_2(\zeta)$ and of the constraints
resulting from $G=\mathcal{N}_{21}=-\mathcal{N}_{12}$. Namely, from the
fact that $f(\zeta)$ is a quotient of polynomials of the same (even)
order, it is clear that the desired two black hole solution can be
generated by a B\"acklund transformation (in our case by a two-fold
B\"acklund transformation) from the Minkowski space. (Note that the axis
values of the Ernst potential determine solutions of the Ernst equation
in a unique way.)

The four constraints $\mathcal{N}_{21}=-\mathcal{N}_{12}$ ensure that the
constants $K_i,\ f_i=-\overline{f}_i\ (i=1,\dots,4)$, $\Omega_0^1,
\Omega_0^2$ may be expressed by two position parameters and the masses
and angular momenta of the two black holes.

The B\"acklund generated solution belonging to \eqref{5.22g} known as the
\lq\lq double Kerr solution\rq\rq\ was intensively discussed by several
authors\cite{19,15}. It turned out that there are necessarily struts between the \lq\lq horizons\rq\rq.
Since we have shown, by solving the boundary value problem, that
B\"acklund generated solutions are the \emph{only} candidates to describe
aligned balanced black holes, we may now assert that black
holes cannot be balanced at all.

\subsubsection{Rigidly rotating disks of dust}

\begin{figure}
\psfrag{z}{$\zeta$}
\psfrag{r}{$\rho$}
\psfrag{disk}{\underline{disk:}}
\psfrag{infinity}{\underline{infinity:}}
\psfrag{f'=e2V}{$f'=\e^{2V_0}$}
\psfrag{fto1}{$f\to 1$}
\psfrag{RK}{$\Re K$}
\psfrag{IK}{$\Im K$}
\psfrag{a}{(a)}
\psfrag{b}{(b)}
\psfrag{r0}{$\rho_0$}
\psfrag{ir0}{$\ii\rho_0$}
\psfrag{-ir0}{$-\ii\rho_0$}
\psfrag{-iz}{$-\ii z$}
\psfrag{+}{\scriptsize{$-$}}
\psfrag{-}{\scriptsize{$+$}}
\psfrag{iz}{$\ii\bar{z}$}
\psfrag{G}{$\Gamma$}
\centerline{\resizebox{12cm}{6cm}{\includegraphics{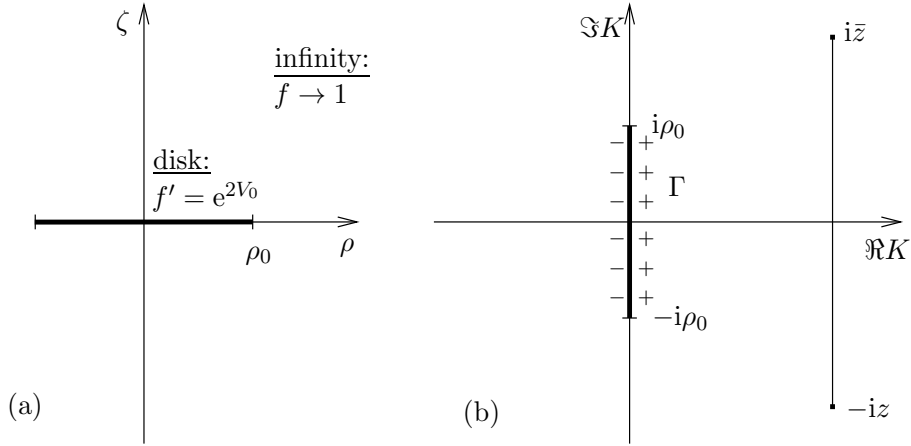}}}
 \caption{(a) Boundary value problem for the rotating disk of dust, (b) 2-sheeted $K$-surface of the rotating disk
 with branch points at $-\ii z, \ii\bar{z}$. \cite{0}}
 \label{fig2}
\end{figure}

Disks of dust can be considered to be extremely flattened spheroids (Fig. 2(a)) consisting of perfect fluid matter \cite{3}. One can 
show that, for rigidly rotating dust, the boundary conditions (\ref{2.4}) and (\ref{2.9}) have to be complemented by the 
condition $b'=\Im f'=0$ on the disk $(\mathcal{B})$. This condition follows from the Einstein equations as a transition condition from a 
divergence-free part of those equations via Gauss's theorem
\cite{3}. Thus we have to take into consideration
\lbe{5.26}
\mathcal{A^\pm}:\quad\text{regularity of}\ f,\qquad\mathcal{B}:\quad f'=\e^{2V_0},\qquad \mathcal{C}:\quad f\to 1,
\ee
see Fig. 2(a). On the disk, the Linear Problem of the co-rotating system of reference takes the form
\lbe{5.27}
\mathcal{B}:\quad \mathbf{\Phi}'_{,\rho}=-\frac{\rho}{\sqrt{K^2+\rho^2}(f'+\overline{f'})}\left(\begin{array}{rr}0&\overline{f'}_{,\zeta}\\[1ex]f'_{,\zeta}&0\end{array}\right)\mathbf{\Phi}',
\ee
where $\mathbf{\Phi}'$ and $f'$ are the \lq co-rotating\rq\
$\mathbf{\Phi}$-matrix and the \lq co-rotating\rq\ Ernst potential on
the disk.
This relation must be discussed under the `boundary conditions'
\begin{equation*}
\mathbf{\Phi}'(\rho=0,\zeta=0^+,\lambda)\big|_\mathcal{B}=\mathbf{\Phi}'(\rho=0,\zeta=0^+,\lambda)\big|_{\mathcal{A}^+},
\end{equation*}
cf. \eqref{5.8} and
\begin{equation*}
\mathbf{\Phi}'(\rho=0,\zeta=0^-,\lambda)\big|_\mathcal{B}=\mathbf{\Phi}'(\rho=0,\zeta=0^-,\lambda)\big|_{\mathcal{A}^-},
\end{equation*}
cf. \eqref{5.9}.

 Again, this discussion allows the construction of $F(K)$ and $G(K)$ and, via 
(\ref{5.12}) and (\ref{5.13}), the construction of the axis values
$f(\zeta)$ of the Ernst potential. We first take advantage of the
symmetry of the problem which implies
$\overline{f(\rho,\zeta)}=f(\rho,-\zeta)$ and connects the
$\zeta$-derivatives of $f'$ above $(\zeta=0^+)$ and below $(\zeta=0^-)$
the disk
\begin{equation}\label{5.27a}
\mathcal{B}:\quad \overline{f'}_{,\zeta}\big|_{\zeta=0^+}=-f'_{,\zeta}\big|_{\zeta=0^-}.
\end{equation}

As a consequence, the LP \eqref{5.27} connects the matrix 
$\stackrel{\text A}{\mathbf{\Phi}}$ above the disk,
\mbox{$\stackrel{\text A}{\mathbf{\Phi}}=\mathbf{\Phi}'(\rho,\zeta=0^+,K)$}, with the matrix 
$\stackrel{\text B}{\mathbf{\Phi}}$ below the disk,
$\stackrel{\text B}{\mathbf{\Phi}}=\mathbf{\Phi}'(\rho,\zeta=0^-,K)$,
\begin{equation}\label{5.27b}
\mathcal{B}:\quad \stackrel{\text A}{\mathbf{\Phi}}=\left(\begin{matrix} 0 & 1 \\ -1 &
0 \end{matrix}\right)\stackrel{\text B}{\mathbf{\Phi}}\mathbf{H}(K),
\end{equation}
where the matrix $\mathbf{H}(K)$ (the \lq\lq integration constant\rq\rq)
does not depend on $\rho\in\mathcal{B}$. At the rim of the disk we have
\begin{equation}\label{5.27c}
\stackrel{\text A}{\mathbf\Phi}(\rho_0,0,K)=
\stackrel{\text B}{\mathbf\Phi}(\rho_0,0,K)=\stackrel{\text r}{\mathbf\Phi}.
\end{equation}
Because of \eqref{5.27b}, the rim matrix \mbox{$\stackrel{\text
r}{\mathbf\Phi}\!\!{}^{-1}\left(\begin{matrix} 0&-1\\1&0\end{matrix}\right)\stackrel{\text
r}{\mathbf\Phi}$} can be expressed in terms of \mbox{$\stackrel{\text
A}{\mathbf{\Phi}}=\mathbf{\Phi}'(\rho,\zeta=0^+,K)$}, $\stackrel{\text
B}{\mathbf{\Phi}}=\mathbf{\Phi}'(\rho,\zeta=0^-,K)$. Note that
$\mathbf\Phi$ is considered to be a holomorphic function of $\lambda$
and therefore a function living on the 2-sheeted Riemann $K$-surface of
Fig. \ref{fig1}(b). Hence we have to discuss the rim matrix as a function
of $K$ on both sheets.

Any $\mathbf\Phi$ multiplied from the right by a matrix function of $K$ is again a
solution of the LP. The discussion of the rim matrix simplifies after
the following redefinition
\begin{equation}\label{5.27d}
\mathcal{R}=\left(\begin{matrix}0&1\\-1&0\end{matrix}\right)
\left(\begin{matrix}F&0\\G&1\end{matrix}\right)
\stackrel{\text r}{\mathbf{\Phi}}\!\!{}^{-1}
\left(\begin{matrix}0&-1\\1&0\end{matrix}\right)
\stackrel{\text r}{\mathbf{\Phi}}
\left(\begin{matrix}F&0\\G&1\end{matrix}\right)^{-1}
\left(\begin{matrix}0&1\\-1&0\end{matrix}\right)^{-1}.
\end{equation}
Using \eqref{5.8}\eqref{5.9} we obtain
\begin{equation}\label{5.27e}
\mathcal{R}=\begin{cases}\e^{2V_0}\mathcal{M}\mathcal{S}^{-1}&\text{on the
upper sheet}\\-\e^{2V_0}\mathcal{S}^{-1}\mathcal{M} & \text{on the lower
sheet}\end{cases}
\end{equation}
where
\begin{equation}\label{5.27f}
\mathcal{M}=\left(\begin{matrix} G(K) & (G^2-1)/F\\-F(K) &
-G(K)\end{matrix}\right),\quad
\mathcal{S}=\left(\begin{matrix} f_0\overline{f_0}-4\Omega_0^2K^2 & \ii
b_0+2\ii\Omega_0 K\\ \ii b_0-2\ii\Omega_0 K & -1\end{matrix}\right)
\end{equation}
and
\begin{equation}\label{5.27g}
f_0=\e^{2V_0}+\ii b_0=f(\zeta=0^+)\ .
\end{equation}
Note that $\mathcal M=\left(\begin{matrix}
0&-1\\1&0\end{matrix}\right)\mathcal N \left(\begin{matrix}
-1&0\\0&1\end{matrix}\right)$.

Obviously, $\tr\mathcal{R}=\tr\mathcal{R}^{-1}=0$ and
$\mathcal{M}^2=\mathbf{1}$, whence
\begin{equation}\label{5.27h}
\tr\mathcal{M}\mathcal{S}^{-1}=\tr\mathcal{SM}=0.
\end{equation}
This relation interlinks $F(K)$ and $G(K)$ and, because of
\eqref{5.12},\eqref{5.13} real and imaginary part of the axis values
$f(\zeta)$ of the Ernst potential\cite{16,NM93}.

We next wish to determine $F(K)$ and $G(K)$ which in turn determine
$f(\zeta)$. To this end we consider $\mathbf\Phi(\rho,\zeta,\lambda)$,
for fixed coordinates $\rho,\zeta$ as a function of $\lambda$. We have
already used the initial conditions $\psi=\chi=1$ for some
$\rho=\rho_0=0$, $\zeta=\zeta_0\in\mathcal{A}^+$ prescribed in one sheet
$(\lambda=-1)$ of the $K$-plane. In principle the behaviour of
$\mathbf\Phi$ in the other sheet and at all points in the $\rho,\zeta$-plane 
can be calculated by integrating the LP along a suitable path. However,
the coefficients $A(\rho,\zeta)$, $B(\rho,\zeta)$ in the LP \eqref{2.14} are not explicitly
known. Nevertheless, their regular behaviour outside the disk together
with the boundary values on the disk, cf. \eqref{5.26}, provides us with
defining properties for $\mathbf\Phi$. One of them may be
taken from Fig. \ref{fig1}(b): Since the domain of the disk,
$0\le\rho\le\rho_0$, $\zeta=0^\pm$ is a non-vacuum domain, where the LP
fails, $\mathbf\Phi$ at the branch point pairs $K=\ii\rho+0^\pm, -\ii\rho+0^\pm$,
$\ 0\le\rho\le\rho_0$ cannot \lq\lq pass\rq\rq\ through the contour $\Gamma:
-\rho_0\le\Im K\le\rho_0$, i.e. $\mathbf\Phi$ has a well-defined jump
between opposite points along the contour $\Gamma$,
see Fig. \ref{fig2}. A careful discussion would show that
$\mathbf\Phi(\rho,\zeta,\lambda)$, for fixed coordinate values
$\rho$,$\zeta$ outside the disk $(\rho,\zeta\notin\mathcal{B})$, is a
regular function in $\lambda$ outside $\Gamma$ and jumps along $\Gamma$,
i.e. $\mathbf\Phi$ satisfies a (regular) Riemann-Hilbert problem.

Consider now the jump $\mathbf\Phi^{-1}_+\mathbf\Phi_-$, where the signs
indicate the two sides of $\Gamma$, cf. Fig. \ref{fig2}(b). The LP tells us that
$\mathbf\Phi^{-1}_+\mathbf\Phi_-$ does not depend on the coordinates and
is therefore a function $\mathcal D$ of the contour alone,
\begin{eqnarray}\label{5.27i}
\mathbf\Phi^{-1}_+\mathbf\Phi_- & =& \mathcal D_\text{u}(K),\qquad
K\in\Gamma_\text{u}\nonumber\\
\mathbf\Phi^{-1}_+\mathbf\Phi_- & =& \mathcal D_\text{l}(K),\qquad
K\in\Gamma_\text{l},
\end{eqnarray}
where u marks the upper and l the lower sheet. Since the jump contours
$\Gamma_\text{u}$, $\Gamma_\text{l}$ and the jump matrices $\mathcal
D_\text{u}$, $\mathcal D_\text{l}$ are the same for all values of
$\rho$, $\zeta$ (i.e. for all Riemann surfaces with different branch
points), we may express $\mathcal D_\text{u}$ and  $\mathcal D_\text{l}$
in terms of the axis values of $\mathbf\Phi$,
\begin{equation}\label{5.27j}
\mathcal{D}_\text{u}(K)=\left(\begin{matrix}F_+ & 0\\G_+ &
1\end{matrix}\right)^{-1}\left(\begin{matrix}F_- & 0\\G_- & 1\end{matrix}\right),
\qquad K\in\Gamma_\text{u}\ .
\end{equation}
A similar relation for $\mathcal D_\text{l}$ may be obtained via
\eqref{5.2a}. As a consequence of \eqref{5.27i}, \eqref{5.27j} the
matrix
$\mathbf\Phi\left(\begin{matrix}F&0\\G&1\end{matrix}\right)^{-1}$ does
not jump along $\Gamma_\text{u}$.
Because of \eqref{5.5} the same holds for
$\mathbf\Phi'\left(\begin{matrix}F&0\\G&1\end{matrix}\right)^{-1}$ and,
finally, for $\mathcal R$ as defined in \eqref{5.27d}. Consider now the
Riemann $K$-surface of the disk rim $\rho=\rho_0$, $\zeta=0$. The cut
between the branch points $K_\text{B}=\pm\ii \rho_0$ coincides with the
contour $\Gamma_\text{u}$, $\Gamma_{l}$ which are on the two \lq\lq
bridges\rq\rq\ connecting crosswise the upper with the lower
sheet. Since $\mathcal{R}$ does not jump on $\Gamma_\text{u}$, we have,
according to \eqref{5.27e}
$(\mathcal{MS}^{-1})_-=-(\mathcal{S}^{-1}\mathcal M)_+$. Though
$\mathcal R$ does not jump, $F$ and $G$ do jump, cf. \eqref{5.27j}. Note
that $F(K)$ and $G(K)$ are unique functions of $K$.
 Hence, there is only one contour $\Gamma:\Re K=0,
-\rho_0\le\Im K\le\rho_0$ where $\mathcal M$ (with the elements $F(K)$,
$G(K)$) does jump. Since $\mathbf\Phi$ is analytic outside
$\Gamma_\text{u},\Gamma_\text{l}$, the matrix $\mathcal M$ must be analytic
outside $\Gamma$. Thus we obtain $F(K)$ and $G(K)$ from the Riemann-Hilbert
problem
\begin{eqnarray}\label{5.27k}
&K\in\Gamma: & \quad\mathcal{SM}_-=-\mathcal M_+\mathcal S\nonumber\\
&K\notin\Gamma: & \quad\mathcal M(K) \text{\ analytic in $K$},
\end{eqnarray}
$\mathcal S$ and $\mathcal M$ as in \eqref{5.27f}.
(Note that the elements of $\mathcal S$, which are polynomials and the
elements of $\mathcal S^{-1}$ which are rational functions in $K$ do not
jump along $\Gamma$.) There is no jump at the end points of the contour
$K=\pm\ii\rho_0$, $\mathcal M(\pm\ii\rho_0)_-=\mathcal
M(\pm\ii\rho_0)_+$. As a consequence, one obtains $\tr\mathcal
S(\pm\ii\rho_0)=0$, i.e., the parameter relation
\begin{equation}\label{5.27kk}
f_0\overline f_0+4\Omega_0^2\rho_0^2=1.
\end{equation}
 It turns out that the Riemann-Hilbert problem (\ref{5.27k}) has a unique
solution $\mathcal M(K)$ in the parameter region
\lbe{5.27l}
0\le \mu=2\Omega_0^2\e^{-2V_0}\rho_0^2<\mu_0=4.62966184\ldots
\ee
An important step on the way to this solution is the diagonalization of
$\mathcal S$. Finally, one obtains $F(K)$, $G(K)$ and the axis values of
the Ernst potential $f(\zeta)$ in terms of elliptic theta functions. We
need not go this road. As we shall see in the next section, we can use
the Riemann-Hilbert problem \eqref{5.27k} to formulate a more general
Riemann-Hilbert problem which will yield the complete disk of dust
solution in terms of hyperelliptic theta functions.

\section{Ernst potential everywhere}
\subsection{Kerr solution}
In the preceding section, we analyzed the axis values of the
Ernst potential. We will now construct the complete solutions
$f(\rho,\zeta)$ of our boundary value problems from the information
about the behaviour along the axis of symmetry gained by the discussion
of the direct problem.

There is, of course, no question that the discussion of the black hole
case in \ref{5.B.1} will lead to the famous Kerr solution (in Weyl
coordinates \eqref{2.1}). The point made here is that this solution
describing the stationary rotating black hole can be derived from a
boundary value problem.

To achieve our goal it is useful to exploit the gauge freedom of
multiplying $\mathbf\Phi$ from the right by an arbitrary matrix funktion
of $K$. The transformation
\begin{equation}\label{5.27m}
\tilde{\mathbf\Phi}=\frac{K^2-\alpha^2-M^2}{K[(K+M)^2+\alpha^2]}\,\mathbf\Phi
\left(\begin{matrix} K+m & \ii\alpha\\ \ii\alpha & K+m\end{matrix}\right)
\end{equation}
preserves the properties \eqref{5.1}--\eqref{5.3} and enables the
calculation of $f$ via \eqref{5.4}. Because of \eqref{5.14},
\eqref{5.222}, \eqref{5.6} and \eqref{2.15}, the determinant of
$\tilde{\mathbf\Phi}$ becomes
\begin{equation}\label{5.27n}
\det\tilde{\mathbf\Phi}=\gamma\frac{(K+\ii
z)^2}{K^2}(\lambda^2-\lambda_1^2)(\lambda^2-\lambda_2^2),
\quad\gamma=\gamma(\rho,\zeta)=-\frac{2\e^{2U(\rho,\zeta)}}{(1-\lambda_1^2)(1-\lambda_2^2)},
\end{equation}
where
\begin{equation}\label{5.27o}
\lambda_i^2=\frac{K_i-\ii \overline z}{K_i+\ii z}\qquad(i=1,2).
\end{equation}
This form of the determinant together with the axis values of
$\tilde{\mathbf\Phi}$ tells us that $\tilde{\mathbf\Phi}$ must be a
quadratic matrix polynomial of $\lambda$,
\begin{equation}\label{5.27p}
\tilde{\mathbf\Phi}=\frac{K+\ii z}{K}(\mathbf{C}+\mathbf{D}\lambda+\mathbf{E}\lambda^2),
\end{equation}
where the $2\times 2$ matrices $\mathbf C$, $\mathbf D$, $\mathbf E$ are
functions of $\rho$, $\zeta$ alone.
It can be shown \cite{13} that $\tilde{\mathbf\Phi}$ with \eqref{5.27n}
satisfies the LP. (It is a B\"acklund transformation of the trivial
solution $f=1$.)

According to \eqref{5.27n}, $\tilde{\mathbf\Phi}(\rho,\zeta,\lambda_i)$
$(i=1,2)$ must have a null eigenvector $b_i$ in the zeros $\lambda_i$,
\begin{equation}\label{5.27q}
\tilde{\mathbf\Phi}(\rho,\zeta,\lambda_i)b_i=0\qquad(i=1,2).
\end{equation}
From the LP it follows that the elements of $b_i$ have to be
constants. Hence, the quotient
\begin{equation}\label{5.27r}
\frac{\tilde\chi(\rho,\zeta,\lambda)}{\tilde\chi(\rho,\zeta,-\lambda)}=-\frac{C_{21}+D_{21}\lambda+E_{21}\lambda^2}{C_{21}-D_{21}\lambda+E_{21}\lambda^2},
\end{equation}
where the coefficients are elements of the matrices $\mathbf C$,
$\mathbf D$, $\mathbf E$ must be a constant at
$\lambda=\lambda_i$ $(i=1,2)$. The values of the two constants $(i=1,2)$ can be read off from the
axis values of $\tilde{\mathbf\Phi}$ resulting from \eqref{5.27r}
together with \eqref{5.6}, \eqref{5.7}, \eqref{5.222}, \eqref{5.223} and \eqref{5.27m},
\begin{equation}\label{5.27s}
\frac{\tilde\chi(\lambda_1)}{\tilde\chi(-\lambda_1)}=-\ii\cot\frac{\varphi}{2},\qquad
\frac{\tilde\chi(\lambda_2)}{\tilde\chi(-\lambda_2)}=\ii\cot\frac{\varphi}{2}.
\end{equation}
Note that $\tilde\chi(-1)=1$ implies
\begin{equation}\label{5.27t}
C_{21}-D_{21}+E_{21}=1.
\end{equation}
These three conditions fix the coefficients $C_{21}$, $D_{21}$, $E_{21}$
via a linear algebraic system. Finally, we obtain the Ernst potential
everywhere from $f=\tilde\chi(1)/\tilde\chi(-1)$,
\begin{equation}\label{5.27u}
f(\rho,\zeta)=\frac{r_1\e^{\ii\varphi}+r_2\e^{\ii\varphi}-2M\cos\varphi}{r_1\e^{\ii\varphi}+r_2\e^{\ii\varphi}+2M\cos\varphi},
\end{equation}
where
\begin{equation*}
r_i^2=(K_i-\zeta)^2+\rho^2\qquad(i=1,2)
\end{equation*}
with $K_1=-K_2$ and $\varphi$ as in \eqref{5.223}. This is the Ernst
potential $f$ of the Kerr solution in Weyl-Papapetrou coordinates. By virtue of \eqref{2.12} and
\eqref{2.13}, this potential determines all metric coefficients in the
line element \eqref{2.1}.

\subsection{Disk of dust solution}
In order to construct the $\mathbf\Phi$-matrix for arbitrary values of
$\rho$, $\zeta$ and $\lambda$, let us return to the Riemann-Hilbert
problem \eqref{5.27k}. As we have seen, the matrix
$\mathbf\Phi\left(\begin{matrix} F&0\\G&1\end{matrix}\right)^{-1}$ does
not jump along $\Gamma_\text{u}$. Analogously,
$\mathbf\Phi\left(\begin{matrix} 1&G\\0&F\end{matrix}\right)^{-1}$ does not
jump along $\Gamma_\text{l}$. The images
$\Gamma_\lambda$ of $\Gamma_\text{u}$ and $\Gamma_{-\lambda}$ of
$\Gamma_\text{l}$ inherit these properties, which are essential to the following deductions.

To formulate a Riemann-Hilbert problem in the $\lambda$-plane, we define
two matrices,
\begin{eqnarray}\label{5.27v}
\mathcal{L} & := & \mathbf\Phi
\left(\begin{matrix}1&G\\0&F\end{matrix}\right)^{-1}
\left(\begin{matrix}1&0\\0&-1\end{matrix}\right)\mathcal M
\left(\begin{matrix}1&0\\0&-1\end{matrix}\right)
\left(\begin{matrix}1&G\\0&F\end{matrix}\right)\mathbf\Phi^{-1}\nonumber\\
& = &
\mathbf\Phi\left(\begin{matrix}F&0\\G&1\end{matrix}\right)^{-1}
\left(\begin{matrix}0&-1\\1&0\end{matrix}\right)\mathcal M
\left(\begin{matrix}0&1\\-1&0\end{matrix}\right)
\left(\begin{matrix}F&0\\G&1\end{matrix}\right)\mathbf\Phi^{-1}\\
& = & \mathbf\Phi\left(\begin{matrix}0&1\\1&0\end{matrix}\right)\mathbf\Phi^{-1},\nonumber
\end{eqnarray}
\begin{eqnarray}\label{5.27w}
\mathcal{Q} & := & \e^{-2V_0}\mathbf\Phi
\left(\begin{matrix}1&G\\0&F\end{matrix}\right)^{-1}
\left(\begin{matrix}1&0\\0&-1\end{matrix}\right)(\mathcal S+w\mathbf 1)
\left(\begin{matrix}1&0\\0&-1\end{matrix}\right)
\left(\begin{matrix}1&G\\0&F\end{matrix}\right)\mathbf\Phi^{-1}\nonumber\\
&=&  \e^{-2V_0}\mathbf\Phi
\left(\begin{matrix}F&0\\G&1\end{matrix}\right)^{-1}
\left(\begin{matrix}0&-1\\1&0\end{matrix}\right)(\mathcal S+w\mathbf 1)
\left(\begin{matrix}0&-1\\1&0\end{matrix}\right)
\left(\begin{matrix}F&0\\G&1\end{matrix}\right)\mathbf\Phi^{-1},
\end{eqnarray}
where
\begin{equation*}
w=-\frac{1}{2}\tr\mathcal S=2\Omega_0^2(K^2+\rho_0^2).
\end{equation*}
Here we have made use of the parameter relation \eqref{5.27kk}.
Since $\mathcal S$ and $w$ are polynomials in $K$ and therefore rational
functions in $\lambda$, the matrix $\mathcal{Q}$ has no jump at all. Taking the
asymptotics of $S$ and $w$ into account, $\mathcal Q$ must take the following
polynomial structure in $\lambda$
\begin{equation}\label{5.27x}
\mathcal{Q}  =  (K+\ii
z)^2\left(\begin{matrix}q_1&q_2\\q_3&-q_2\end{matrix}\right),
\ q_1=k\lambda+l\lambda^3,\ q_2=m+n\lambda^2+p\lambda^4,\ q_3=q+r\lambda^2+s\lambda^4,
\end{equation}
where $k,l,m,n,p;q,r,s$ are functions of $\rho$, $\zeta$ alone. From the
definitions \eqref{5.27v}, \eqref{5.27w} and the condition
\eqref{5.27h}, we may derive
\begin{equation}\label{5.27y}
\mathcal{QL}=-\mathcal{LQ}
\end{equation}
whereas the particular Riemann-Hilbert problem \eqref{5.27k} has the
continuation
\begin{equation}\label{5.27z}
\begin{array}{lc}
\lambda\in\Gamma_\lambda: & (\mathcal Q+\e^{-2V_0}w\mathbf 1)\mathcal
L_-=-\mathcal L_+(\mathcal Q+\e^{-2V_0}w\mathbf 1)\\
\lambda\in\Gamma_{-\lambda}: & (\mathcal Q-\e^{-2V_0}w\mathbf 1)\mathcal
L_-=-\mathcal L_+(\mathcal Q-\e^{-2V_0}w\mathbf 1)\\
\lambda\notin\Gamma_\lambda,\Gamma_{-\lambda}: & \mathcal L\text{\
analytic in $\lambda$}
\end{array}
\end{equation}

The following solution of the regular Riemann-Hilbert problem
\eqref{5.27z} is based on the diagonalization of $\mathcal Q$.


We consider a function $\Psi$ defined by
\begin{equation}
\Psi:=\frac{1}{\sqrt{w^2+\e^{4V_0}}}\,\ln\frac{\hat{\mathcal
L}_{22}+\sqrt{1+w^2\e^{-4V_0}}\,\hat{\mathcal L}_{21}}
      {\hat{\mathcal L}_{22}-\sqrt{1+w^2\e^{-4V_0}}\hat{\mathcal L}_{21}},
\end{equation}
where
\begin{equation}
\hat{\mathcal L}=\mathcal L\left(\begin{matrix} 1&\mathcal
Q_{11}\\0&\mathcal Q_{21}\end{matrix}\right).
\end{equation}
Note that $\Psi$ has no branch points at the zeroes $K_1$, $K_2$, $\overline
K_1=-K_2$ and $\overline K_2=-K_1$ of $w^2+\e^{4V_0}$,
\begin{equation}\label{5.28}
K_1^2=\rho_0^2\frac{\ii-\mu}{\mu},\quad
K_2^2=\rho_0^2\frac{\ii+\mu}{\mu}\quad(\Re K_1<0,\quad \Re
K_2>0,\quad\mu\ \text{as in \eqref{5.27l}})
\end{equation}
($\Psi$ is unaffected by a change in the sign of
$\sqrt{w^2+\e^{4V_0}}$). It
is an odd function of $\lambda$ (vanishing at $\lambda=0$ and at $\lambda=\infty$). Therefore,
the function
\begin{equation}
\hat{\Psi}=\Psi/[\lambda(K+\ii z)]=\Psi/\sqrt{(K-\ii\bar{z})(K+\ii z)}
\end{equation}
can be discussed as a unique function of $K$
with the following properties:

\medskip

\noindent (i) Along $\Gamma$, because of \eqref{5.27z}, it jumps according to
\begin{equation}
\hat{\Psi}_-=\hat{\Psi}_++\frac{2}{\sqrt{(K-\ii\bar{z})(K+\ii z)}\sqrt{w^2+\e^{4V_0}}}
\,\ln\frac{\sqrt{w^2+\e^{4V_0}}+w}{\sqrt{w^2+\e^{4V_0}}-w}, .
\end{equation}
(ii) Because of
\begin{equation}
\hat{\mathcal L}^2_{21}(1+w^2\e^{-4V_0})-\hat{\mathcal L}^2_{22}=\mathcal Q^2_{21},
\end{equation}
\begin{equation}
\mathcal Q_{21}=-\frac{2f\Omega_0^2\e^{-2V_0}}{f+\overline f}(K-K_a)(K-K_b),
\end{equation}
the behaviour for $K\to K_{a/b}$ is given by
\begin{equation}
\hat{\Psi}\to \frac{\pm 2}{\sqrt{(K_{a/b}-\ii\bar{z})(K_{a/b}+\ii z)(w_{a/b}^2+\e^{4V_0})}}
   \,\ln (K-K_{a/b}) \quad \mbox{as} \quad  K\to K_{a/b}.
\end{equation}
(The ambiguity of sign can be compensated for by the square root.)\\
(iii) The behaviour for $K\to\infty$, because of the definitions of
$\mathcal Q$ and $\mathcal L$, is given by
\begin{equation}
\hat{\Psi}\to \frac{\ln f}{\Omega^2K^3} \quad \mbox{as} \quad K\to\infty.
\label{as}
\end{equation}

These properties are realized
by the following representation of $\hat{\Psi}$:
\begin{eqnarray}
\hat{\Psi}=\frac{1}{\pi\ii}\int\limits_{-\ii\rho_0}^{\ii\rho_0}
     \frac{\ln\frac{\sqrt{w'^2+\e^{4V_0}}+w'}{\sqrt{w'^2+\e^{4V_0}}-w'}}
     {\sqrt{(K'-\ii\bar{z})(K'+\ii z)}\sqrt{w'^2+\e^{4V_0}}(K'-K)}\,\dd K' \nonumber \\
     -\,2\int\limits_{K_1}^{K_a}\frac{1}
     {\sqrt{(K'-\ii\bar{z})(K'+\ii z)(w'^2+\e^{4V_0})}(K'-K)}\,\dd K' \nonumber \\
     -\,2\int\limits_{K_2}^{K_b}\frac{1}
     {\sqrt{(K'-\ii\bar{z})(K'+\ii z)(w'^2+\e^{4V_0})}(K'-K)}\,\dd K',
\label{Psi}
\end{eqnarray}
where $K_a$ and $K_b$ have to be determined such that $\hat{\Psi}={\cal O}(K^{-3})$.
The lower limits of integration  in the last two integrals have been fixed to obtain the
correct result in the Newtonian limit $\mu\to 0$ where $K_a/K_1=1+{\cal O}(\mu^2)$ and
$K_b/K_2=1+{\cal O}(\mu^2)$. (A systematic post-Newtonian expansion of the solution is given
in \cite{PM01}.) Note that the last two terms in Eq.~(\ref{Psi}) may also be interpreted
as follows,
\begin{equation}
2\,(\int\limits_{K_1}^{K_a} + \int\limits_{K_2}^{K_b})=
2\,(\int\limits_{K_a}^{K_1}\{-\} + \int\limits_{K_2}^{K_b})=
\int\limits_{K_a}^{K_b}\{1\} + \int\limits_{K_a}^{K_b}\{2\},
\end{equation}
showing that nothing special happens at $K_1$ and $K_2$.
In this symbolic notation $\{-\}$ indicates that the square root is meant to have the
opposite sign with reference to the first term; $\{1\}$
and $\{2\}$ denote
different paths in the complex $K$-plane, which are chosen such that the closed integral
\begin{equation}
\oint=\int\limits_{K_a}^{K_b}\{1\} - \int\limits_{K_a}^{K_b}\{2\}=2\,\int\limits_{K_1}^{K_2}
\end{equation}
is performed around a contour enclosing the branch points $K_1$ and $K_2$ of
$\sqrt{w^2+\e^{4V_0}}$. In the subsequent formulae we normalize $K$  and introduce
\begin{equation}
X=\frac{K}{\rho_0},\quad X_{a/b}=\frac{K_{a/b}}{\rho_0},\quad X_{1/2}=\frac{K_{1/2}}{\rho_0}.
\end{equation}

An asymptotic expansion of Eq.~(\ref{Psi}) for $X\to\infty$ ($K\to\infty$) leads, according to (\ref{as}), to


\begin{equation}
\ln f = 
\mu\left[\,\int\limits_{X_1}^{X_a}\frac{X^2\,\dd X}{W} +  
\int\limits_{X_2}^{X_b}\frac{X^2\,\dd X}{W} - 
\int\limits_{-\ii}^{\ii}\frac{hX^2\,\dd X}{W_1}\right],
\label{fint}
\end{equation}

\begin{equation}
\int\limits_{X_1}^{X_a}\frac{\dd X}{W} +
\int\limits_{X_2}^{X_b}\frac{\dd X}{W} = 
\int\limits_{-\ii}^\ii \frac{h \dd X}{W_1},
\qquad \int\limits_{X_1}^{X_a}\frac{X\,\dd X}{W} + 
\int\limits_{X_2}^{X_b}\frac{X\,\dd X}{W} = 
\int\limits_{-\ii}^\ii \frac{h X \dd X}{W_1},
\label{jacobi}
\end{equation}
where the lower integration limits $X_1$, $X_2$ are given by
\begin{equation}
X_1^2 = \frac{\ii - \mu}{\mu}, \quad  X_2^2 = -\frac{\ii + \mu}{\mu}
\quad (\Re X_1 < 0, \quad \Re X_2 > 0),
\end{equation}
whereas the upper limits $X_a$, $X_b$ must be calculated from the
integral equations \eqref{jacobi}.
\noindent
Here we have introduced the abbreviations
\begin{equation}\label{105}
W = W_1 W_2, \qquad W_1 = \sqrt{(X-\zeta/\rho_0)^2 +
                              (\rho/\rho_0)^2}, \qquad
W_2 = \sqrt{1 + \mu^2(1+X^2)^2}
\end{equation}
and
\begin{equation}
h = \frac{\ln\left(\sqrt{1 + \mu^2(1+X^2)^2} + \mu(1+X^2)\right)}
         {\pi\ii \sqrt{1 + \mu^2(1+X^2)^2} } \,.
\end{equation}
The third integral in (\ref{fint}) as well as the integrals on the 
right-hand sides in (\ref{jacobi}) have to be taken 
along the imaginary axis in the complex X-plane
with $h$ and and $W_1$ fixed according to $\Re W_1<0$ (for $\rho,\zeta$
outside the disk) and $\Re h = 0$\,.
The task of calculating the upper limits $X_a$, $X_b$ in \eqref{jacobi} from
\begin{equation}
u = \int\limits_{-\ii}^\ii \frac{h\, dX}{W_1} \,, \quad
v = \int\limits_{-\ii}^\ii \frac{h X dX}{W_1}
\label{uv}
\end{equation}
is known as Jacobi's inversion problem. 
G\"opel \cite{goe} and Rosenhain
\cite{ro} were able to express the hyperelliptic functions $X_a(u,v)$ and
$X_b(u,v)$ in terms of (hyperelliptic) theta functions. Later on it turned
out that even the first two integrals in (\ref{fint}) can be expressed by
theta functions in $u$ and $v$! A detailed introduction
into the related mathematical theory which was founded by Riemann  
and Weierstra\ss\  may be found in \cite{st,kra,bob}.
The representation of the Ernst potential (\ref{fint}) 
in terms of theta functions can
be be taken from Stahl's book, see \cite{st}, page 311, Eq.~(5). Here is the
result: Defining a theta function
$\vartheta(x,y;p,q,\alpha)$
by
\begin{equation}
\vartheta(x,y;p,q,\alpha) = \sum\limits_{m=-\infty}^{\infty}
                         \sum\limits_{n=-\infty}^{\infty}
(-1)^{m+n} p^{m^2} q^{n^2} \e^{2mx + 2ny + 4m n \alpha}
\label{theta}
\end{equation}
one can reformulate the expressions (\ref{fint}), (\ref{jacobi}) to give
\begin{equation}
f = \frac{ \vartheta(\alpha_0 u + \alpha_1 v - C_1,
                  \beta_0 u + \beta_1 v - C_2;     p,q,\alpha) }
         { \vartheta(\alpha_0 u + \alpha_1 v + C_1,
                  \beta_0 u + \beta_1 v + C_2;     p,q,\alpha) } \,\,
    \e^{-(\gamma_0 u + \gamma_1 v + \mu w)}
\label{ftheta}
\end{equation}
with $u$ and $v$ as in (\ref{uv}) and
\begin{equation}
w = \int\limits_{-\ii}^\ii \frac{h X^2 \dd X}{W_1} \,.
\end{equation}

The normalization parameters $\alpha_0$, $\alpha_1$; $\beta_0$, $\beta_1$;
$\gamma_0$, $\gamma_1$, the moduli $p$, $q$, $\alpha$
of the theta function and the quantities $C_1$, $C_2$ are defined on the two
sheets of the hyperelliptic Riemann surface related to
\begin{equation}
W = \mu \sqrt{(X-X_1)(X-\bar{X_1})(X-X_2)(X-\bar{X_2})
              (X-\ii\bar{z}/\rho_0)(X+\ii z/\rho_0)},
\end{equation} 
see Figure \ref{riemann}. 
\begin{figure}
\includegraphics[width=0.7\textwidth]{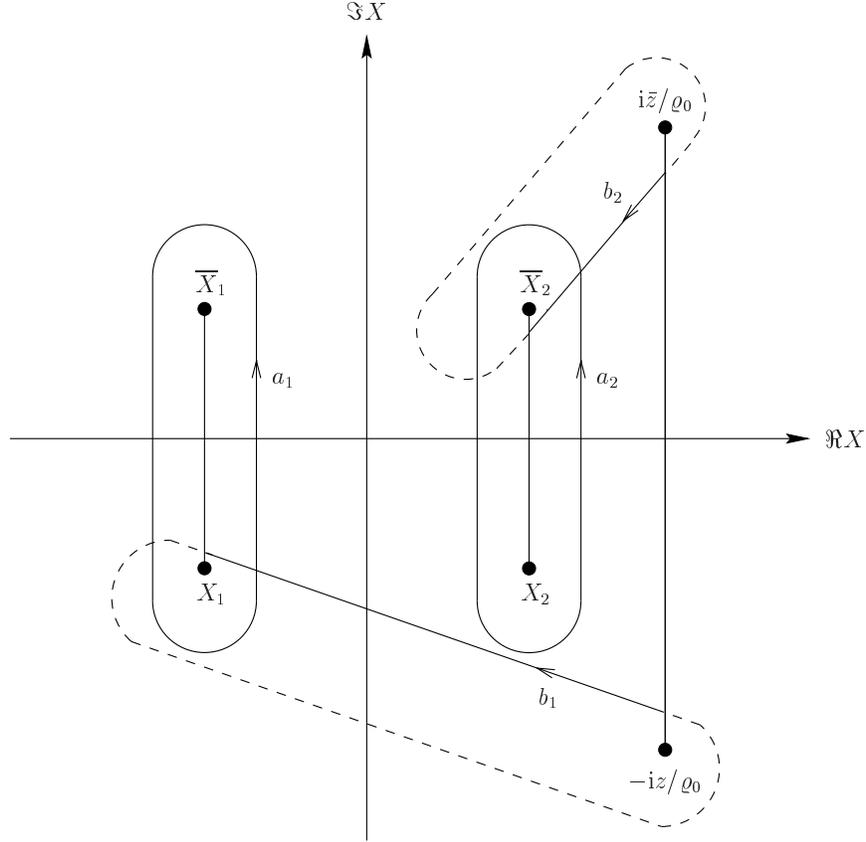}
\caption{Riemann surface with cuts between the 
branch points $X_1$ and
$\bar{X}_1$, $X_2$ and $\bar{X}_2$, $-\ii z/\rho_0$ and $\ii\bar{z}/\rho_0$.
Also shown are the four periods $a_i$ and $b_i$ ($i=1,2$). 
(Continuous/dashed lines 
belong to the upper/lower sheet defined by $W\rightarrow\pm\mu X^3$ as 
$X\rightarrow\infty$.) \cite{3}}
\label{riemann}
\end{figure}
There are two normalized Abelian differentials
of the first kind
\begin{eqnarray}
\dd\omega_1 & = & \alpha_0 \frac{\dd X}{W} + \alpha_1 \frac{X \dd X}{W} \\
\dd\omega_2 & = & \beta_0 \frac{\dd X}{W} + \beta_1 \frac{X \dd X}{W}
\end{eqnarray}
defined by
\begin{equation}
\oint\limits_{a_m} \dd\omega_n = \pi\ii\, \delta_{mn} \, \quad (m=1,2;\, 
n=1,2)\,.
\label{aper}
\end{equation}
Eq.~(\ref{aper}) consists of four linear algebraic equations 
and yields the four
parameters $\alpha_0$, $\alpha_1$, $\beta_0$, $\beta_1$ in terms of integrals
extending over the closed (deformable) curves $a_1$, $a_2$. It can be shown
that there is one normalized Abelian differential of the third kind
\begin{equation}
\dd\omega = \gamma_0 \frac{\dd X}{W} + \gamma_1 \frac{X \dd X}{W}
              + \mu \frac{X^2 \dd X}{W}
\end{equation}
with vanishing $a-$periods,
\begin{equation}
\oint\limits_{a_j} \dd\omega = 0 \,\quad (j=1,2)\,.
\end{equation}
This equation defines $\gamma_0$, $\gamma_1$ (again via a linear algebraic
system). The Riemann matrix
\begin{equation}
(B_{i j}) = \left( \begin{array}{cc}
                     \ln p & 2\alpha \\
                     2\alpha & \ln q
                   \end{array} \right) \, \quad (i=1,2;\, j=1,2)
\end{equation}
(with negative definite real part) is given by
\begin{equation}
B_{i j} = \oint\limits_{b_i} d\omega_j
\end{equation}
and defines the moduli $p$, $q$, $\alpha$ of the theta function (\ref{theta}).
Finally, the quantities $C_1$, $C_2$ can be calculated by
\begin{equation}
C_i = -\int\limits_{-\ii z/\rho_0}^{\infty^+} \dd\omega_i \, \quad (i=1,2) \,,
\end{equation}
where $+$ denotes the upper sheet.
Obviously, all the quantities entering the theta functions and the
exponential function in (\ref{ftheta}) 
can be expressed in terms of well--defined
integrals and depend on the three parameters $\rho/\rho_0$,
$\zeta/\rho_0$, $\mu$.
The corresponding ``tables'' for $\alpha_i$, $\beta_i$, $\gamma_i$, 
$C_i$, $B_{i j}$,
$u$, $v$, $w$ can easily be calculated by numerical integrations. Fortunately,
theta series like (\ref{theta}) converge rapidly. For $0<\mu<\mu_0$,
the solution (\ref{ftheta}) is analytic {\it everywhere} outside the disk ---
even at the rings $-\ii z/\rho_0=X_1$, $X_2$. The complete metric,
calculated according to \eqref{2.1} and \eqref{2.11a}--\eqref{2.13} is
given in the Appendix.

In the framework of the completely integrable evolution equations, the
solution \eqref{ftheta}
may be interpreted as a `B\"acklund-like' transformation of well-defined `seed' solutions $u, v, w$
satisfying axisymmetric Laplace equations. The transformation `parameters' $\alpha_0, \beta_0;
\alpha_1, \beta_1;\gamma_0, \gamma_1; p,q, \alpha; C_1, C_2$ depend on the 6 branch points of the 2-sheeted Riemann
$K$-surface associated with the function $W = W(X)$, cf. \eqref{105},
and do not depend on $u, v, w$. All in all, $f$ is a function of the 2 parameters
$\rho_0$ and $\mu$ and the 2
cylindrical coordinates $\rho$ and $\zeta$. For $\mu \ll 1$ we obtain the
Maclaurin disk as the Newtonian limit.

\section{Physical discussion}

\begin{figure}
\includegraphics{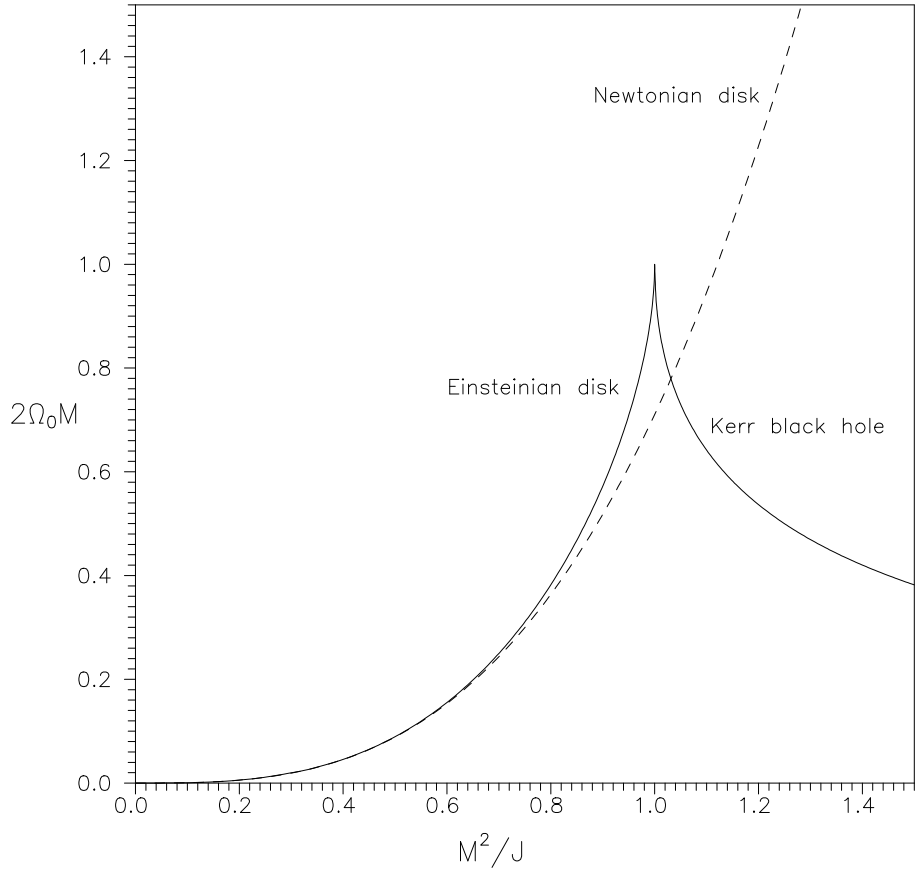}
\caption{Relation between $2\Omega_0 M$ and $M^2/J$ for the classical
Maclaurin disk (dashed line), the general-relativistic dust disk and the
Kerr black-hole \cite{NM93}}
\label{fig5}
\end{figure}
\begin{figure}
\centerline{(a)\resizebox{5.5cm}{5.5cm}{\includegraphics{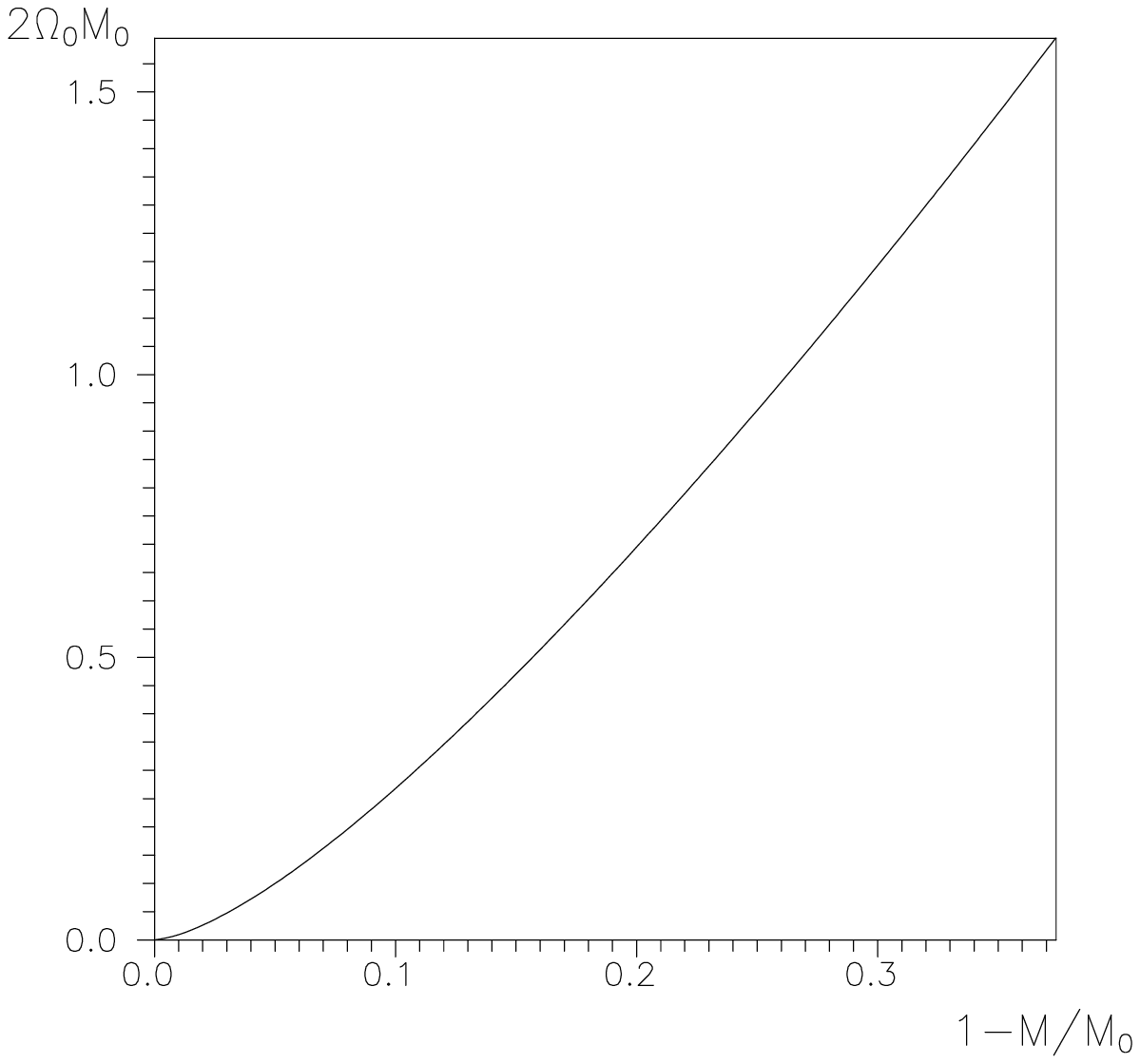}}\hspace*{0.7cm}(b)\resizebox{5.5cm}{5.5cm}{\includegraphics{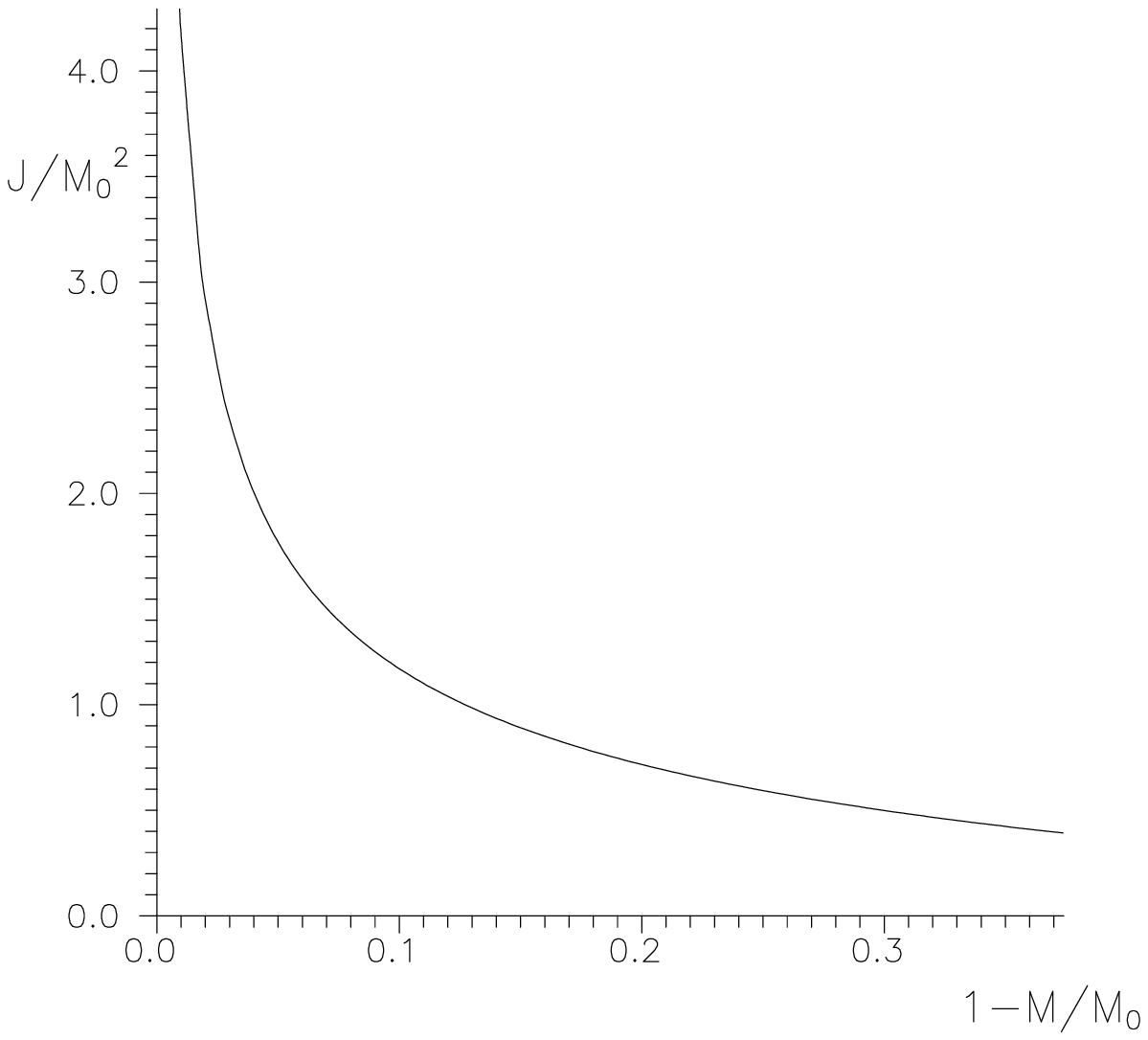}}}
 \caption{(a) $\Omega_0 M_0$ and (b) $J/M_0^2$ as functions of the
 relative binding energy $(M_0-M)/M_0$ for the disk of dust. \cite{0}}
 \label{f12}
\end{figure}

Since the Kerr black hole is also a 2 parameter solution it might be
interesting to compare the behaviour of both solutions in dependence on
common parameters, say, on mass $M$ and angular momentum $J$. It must be
possible to express the area of the horizon and the disk, the radius
$\rho_0$ of the disk or other physical quantities in terms of $M$ and
$J$. A very illustrative relation is the angular velocity $\Omega_0$ as
a function of $M$ and $J$, since $\Omega_0$ is defined in both
cases. For black holes, we have derived the explicit expression
\eqref{5.225}. Surprisingly, the corresponding disk of dust relation has
the same scaling behaviour, i.e., $M\Omega_0$ is a function of $M^2/J$ alone.
Fig. \ref{fig5} shows this dependence
for both solutions \cite{NM93}.
For $M^2/J\to1$ (corresponding to $\mu\to\mu_0$), where the disk solution becomes identical with
the extreme Kerr solution outside the horizon ($\rho^2+\zeta^2>0$),
there is a \lq\lq phase transition\rq\rq\ between the disk and the black
hole. Note that for non-vanishing
$\Omega_0$, $\rho_0\to 0$ as $\mu\to\mu_0$. A detailed analysis of the disk solution for
$\mu\to\mu_0$, including the discussion of a different, non-asymptotically flat limit of
space-time, which is obtained for finite $\rho/\rho_0$ and $\zeta/\rho_0$
($\rho^2+\zeta^2=0$\quad !),
can be found in \cite{M02}.

 We remark that (\ref{ftheta}) solves the Bardeen-Wagoner problem \cite{17} explicitly. All metric
coefficients in (1) are analytic in $\rho, \zeta$ outside the disk and continuous through the
disk. From a physical point of view we have an extremely flattened rigidly rotating body and,
likewise, a rotating continuous distribution of mass points interacting via gravitational forces
alone (`galaxy' model). Fig.~\ref{f12} illustrates the `parametric' collapse of a disk with the total mass-energy $M$,
the baryonic mass $M_0$, the angular velocity $\Omega_0$ and the angular momentum $J$ towards the
black hole limit $(1 - M/M_0 = 0.3732835\ldots)$. Imagine a disk consisting of a fixed number
of baryons (fixed $M_0$): Occupying states with decreasing energy $M$, it would
shrink thereby shedding angular momentum but increasing its angular velocity. The above mentioned
limit of the relative binding energy 1 - $M$/$M_0$ corresponds to the extreme
black hole limit. Additional physical effects (ergozones, dragging effects, surface mass density
\ldots) as well as further parameter relations  have been discussed in \cite{3} and \cite{18}. 

\bigskip

The methods outlined in this paper could be used to construct self-gravitating disks around a
central black hole.

\appendix*
\section{}
The metric functions $\e^{2U}$, $a$, $\e^{2k}$ calcutated from the
Ernst potential \eqref{ftheta} via \eqref{2.11a}--\eqref{2.13}
are given as follows:
\begin{equation*}
\e^{2U}=\frac{\vartheta(\mathbf c)\vartheta^*(\mathbf
c)\vartheta(\mathbf a)\vartheta^*(\mathbf a)}{\vartheta(\mathbf 0)\vartheta^*(\mathbf
0)\vartheta(\mathbf a+\mathbf c)\vartheta^*(\mathbf a+\mathbf c)}
\e^{-(\gamma_0u+\gamma_1 v+\mu w)},
\end{equation*}
\begin{equation*}
1+\frac{(1+\Omega_0 a)\e^{2U}}{\Omega_0\rho}=\frac{\vartheta(\mathbf 0)\vartheta^*(\mathbf
0)\vartheta(\mathbf a+2\mathbf c)\vartheta^*(\mathbf a)}{\vartheta(\mathbf c)\vartheta^*(\mathbf
0)\vartheta(\mathbf a+\mathbf c)\vartheta^*(\mathbf a+\mathbf c)},
\end{equation*}
\begin{equation*}
\e^{2k(\rho,\zeta)}=\frac{\kappa(\rho,\zeta)}{\kappa(0,0)}
\end{equation*}
with
\begin{equation*}
\kappa(\rho,\zeta)=\frac{\vartheta(\mathbf a)\vartheta^*(\mathbf
a)}{\vartheta(\mathbf 0)\vartheta^*(\mathbf
0)}\exp\left({2k_0-\frac{1}{2}}\sum\limits_{i,k=1}^2 a_i
a_k\frac{\partial^2\ln\vartheta(\mathbf x)\vartheta^*(\mathbf
x)}{\partial x_i\partial x_k}\bigg|_{\mathbf x=\mathbf 0}\right),
\end{equation*}
where
\begin{equation*}
2k_0=\frac{\mu^2}{4}\int\limits_{-\ii}^\ii\int\limits_{-\ii}^\ii\dd
X\dd X'\frac{(\lambda-\lambda')^2}{\lambda\lambda'}\,\frac{h(X)h(X')(X-X_1)(X-X_2)(X'+X_1)(X'+X_2)}{(X-X')^2},
\end{equation*}
\begin{equation*}
\lambda=\sqrt{\frac{X-\ii\overline z/\rho_0}{X+\ii z/\rho_0}},\qquad
\lambda'=\sqrt{\frac{X'-\ii\overline z/\rho_0}{X'+\ii z/\rho_0}},
\end{equation*}
\begin{equation*}
\vartheta(\mathbf x)=\vartheta(\mathbf x;p,q,\alpha)=\vartheta(x_1,x_2;p,q,\alpha),
\end{equation*}
\begin{equation*}
\vartheta^*(\mathbf x)=\vartheta(x_1+\frac{\ii\pi}{2},x_2+\frac{\ii\pi}{2};p,q,\alpha),
\end{equation*}
\begin{equation*}
\mathbf a=(a_1,a_2)=(\alpha_0 u+\alpha_1 v, \beta_0 u+\beta_1 v),\quad
\mathbf 0=(0,0),\quad \mathbf c=(C_1,C_2).
\end{equation*}
\begin{acknowledgments}
We wish to thank
Andreas Kleinw\"achter for numerous discussions and J\"org Hennig for technical assistance.
\end{acknowledgments}



\begin{thebibliography}{99}
\bibitem{than} A. S. Fokas, L.-Y. Sung, D. Tsoubelis, Mathematical
Physics, Analysis and Geometry \textbf{1} (1999) 313.
\bibitem{0} G. Neugebauer, Ann. Phys. (Leipzig) \textbf{9} (2000) 342.
\bibitem{3} G. Neugebauer, A. Kleinw\"achter, and R. Meinel, Helv. Phys. Acta \textbf{69} (1996) 472.
\bibitem{4} D. Maison, Phys. Rev. Lett. \textbf{41} (1978) 521.
\bibitem{5} V. A. Belinski and V. E. Zakharov, Zh. Eksper. Teoret. Fiz. Pis'ma \textbf{75} (1978) 195.
\bibitem{6} B. K. Harrison, Phys. Rev. Lett. \textbf{41} (1978) 119.
\bibitem{7} G. Neugebauer, J. Phys. \textbf{A} \textbf{12} (1978) L67; \textbf{A 13} (1980) L19.
\bibitem{8} I. Hauser and F. J. Ernst, Phys. Rev. D \textbf{20} (1979) 362 and 1783; J. Math. Phys. \textbf{21}               
            (1980) 1418.
\bibitem{9} C. Hoenselaers, W. Kinnersley, and B. C. Xantopoulos, Phys. Rev. Lett. \textbf{42} (1979) 481.
\bibitem{10} G. Neugebauer and D. Kramer, J. Phys. \textbf{A} \textbf{16} (1983) 1927.
\bibitem{11} R. Meinel and G. Neugebauer, Class. Quantum Grav. \textbf{12} (1995) 2045.
\bibitem{12} B. Carter, {\em Black Hole Equilibrium States}, in {\em Black Holes} edited by C. and B. de Witt, 
Gordon and Breach, New York 1973.
\bibitem{14} G. Krenzer, G. Neugebauer, and R. Meinel, to be published.
\bibitem{19} D. Kramer and G. Neugebauer, Phys. Lett. \textbf{A75}
(1980) 259.
\bibitem{15} W. Dietz and C. Hoenselaers, Ann. Phys. \textbf{165} (1985) 319;\newline
A. Tomimatsu and M. Kihara, Progr. Theor. Phys. \textbf{67} (1982) 349 and 1406; \newline
D. Kramer, Gen. Rel. and Grav. (GRG) \textbf{18} (1986) 497;\\
V. S. Manko and E. Ruiz, Class. Quantum Grav. {\bf 18} (2001) L11.
\bibitem{16} G. Neugebauer and R. Meinel, Phys. Rev. Lett. \textbf{73} (1994) 2166, Phys. Rev. Lett. \textbf{75} (1995) 3046.
\bibitem{NM93} G. Neugebauer and R. Meinel, Astrophys. J. {\bf 414} (1993) L97.
\bibitem{13} G. Neugebauer, {\em Gravitostatics and Rotating Bodies}, in {\em General Relativity} edited by G.  
S. Hall and J. R. Pulham, SUSSP Publications and Institute of Physics Publishing, Edinburgh University and The Institute of Physics, London 1996.

\bibitem{PM01}  D. Petroff and R. Meinel, Phys. Rev. D {\bf 63} (2001) 064012.
\bibitem{goe} A.~G\"opel, Crelle's J.~f\"ur Math.~{\bf 35} (1847) 277.
\bibitem{ro} G.~Rosenhain, Crelle's J.~f\"ur Math.~{\bf 40} (1850) 319.
\bibitem{st} H.~Stahl, {\it Theorie der Abel'schen Funktionen} (Teubner 1896).
\bibitem{kra} A.~Krazer, {\it Lehrbuch der Thetafunktionen} (Teubner 1903).
\bibitem{bob} E.D.~Belokolos, A.I.~Bobenko, V.Z.~Enol'skii, A.R.~Its and
V.B.~Matveev, {\it Algebro-Geometric Approach to Nonlinear Integrable
Equations} (Springer 1994).
\bibitem{M02} R. Meinel, Ann. Phys. (Leipzig) {\bf 11} (2002) 509 [gr-qc/0205127].
\bibitem{17} J. M. Bardeen and R. V. Wagoner, Astrophys. J. \textbf{158} (1969) L65; \textbf{167}
(1971) 359.
\bibitem{18} R. Meinel and A. Kleinw\"achter, Einstein Studies (Birkh\"auser) \textbf{6} (1995)
339.

\end{thebibliography}
\end{document}